\documentclass[useAMS,onecolumn]{mn2e}

\newif\ifAMStwofonts

\usepackage{graphicx}	
\usepackage{amssymb}	
\usepackage{epsfig}
\usepackage{color}

\def\pmb#1{\mbox{\boldmath$#1$}}
\def\gtsim {>\kern-1.2em\lower1.1ex\hbox{$\sim$}}
\def\ltsim {<\kern-1.2em\lower1.1ex\hbox{$\sim$}}
\def\gtsim {>\kern-1.2em\lower1.1ex\hbox{$\sim$}}
\def\ltsim {<\kern-1.2em\lower1.1ex\hbox{$\sim$}}

\def\be{\begin{equation}}
\def\ee{\end{equation}}

\def\be{\begin{eqnarray}}
\def\ee{\end{eqnarray}}

\def\pmbmt#1{\pmb{\sf #1}}
\def\rmi{{\rm i}}

\begin{document}

\title[Tidal Oscillations of Hot Jupiters]{Tidal Oscillations of Rotating Hot Jupiters}

\author[U. Lee]{
Umin Lee$^{1}$\thanks{E-mail: lee@astr.tohoku.ac.jp}
\\
$^{1}$Astronomical Institute, Tohoku University, Sendai, Miyagi 980-8578, Japan\\
}

\date{Accepted XXX. Received YYY; in original form ZZZ}
\pubyear{2015}

\maketitle

\begin{abstract}
We calculate small amplitude gravitational and thermal tides of 
uniformly rotating hot Jupiters composed of
a nearly isentropic convective core and a geometrically thin radiative envelope.  
We treat the fluid in the convective core as a viscous fluid and
solve linearized Navier Stokes equations to obtain tidal responses of the core, 
assuming that the Ekman number ${\rm Ek}$ is a constant parameter.
In the radiative envelope, we take account of the effects of 
radiative dissipations on the responses.
The properties of tidal responses depend on thermal timescales $\tau_*$ in the envelope and
Ekman number Ek in the core and on weather the forcing frequency $\omega$ is in the inertial range or not, where
the inertial range is defined by $|\omega|\le2\Omega$ for the rotation frequency $\Omega$.
If ${\rm Ek}\gtrsim 10^{-7}$, the viscous dissipation in the core is dominating the thermal contributions
in the envelope for $\tau_*\gtrsim 1$ day.
If ${\rm Ek}\lesssim 10^{-7}$, however, the viscous dissipation is comparable to or smaller than
the thermal contributions and the envelope plays an important role to determine the tidal torques.
If the forcing is in the inertial range, frequency resonance of the tidal forcing with core inertial modes 
significantly affects the tidal torques, producing numerous resonance peaks of the torque.
Depending on the sign of the torque in the peaks, we suggest that there exist cases in which
the resonance with core inertial modes hampers the process of synchronization between the spin and orbital motion
of the planets.
\end{abstract}



\section{introduction}

Hot Jupiters are giant gas planets with orbital periods of a few days and
most of them have relatively small eccentricities of the orbits (e.g., Udry \& Santos 2007).
Because of their proximity to the host stars, tides on the planets (and the stars) are believed to
play important roles both in the processes of their formation and in determining
their observed properties.

Origins of hot Jupiters, however, are not necessarily well understood (e.g., Dawson \& Johnson 2018; Stevenson 1982).
Some scenarios proposed for the origins
assume the existence of epochs in which the tidal effects become crucial.
In high-eccentricity tidal migration scenario, for example, 
dissipative tidal interactions between the planets and stars
are essential to the evolution of the orbits. 
Assuming that the hot Jupiter HD80606b was produced through 
high-eccentricity tidal migration, Wu \& Murray (2003) numerically followed the orbit evolution of a Jovian planet
around the hosting star in the binary system of stars, where they employed
the formulation by Eggleton, Kiseleva, \& Hut (1998) and Eggleton \& Kiseleva-Eggleton (2001),
which takes account of tidal frictions between the planets and stars. 
See, e.g., Ogilvie (2014) for a review of tides in stars and planets.

Many hot Jupiters are known to have radii larger than those of Jovian planets orbiting far from the host stars 
(e.g., Dawson \& Johnson 2018; Jermyn et al 2017).
Calculating cooling evolution of Jovian planets, 
Baraffe et al (2003) suggested that additional heating sources in the interior could be effective to inflate the radii of hot Jupiters. 
Several authors have examined as such heating sources tidal mechanisms, that is,
(gravitational) tidal heating (e.g., Bodenheimer et al 2001) and
thermal tides (e.g., Arras \& Socrates 2010; Socrates 2013, Auclair-Desrotour \& Leconte 2018).
See also, e.g., Jermyn et al (2017) and Lee (2019) for other possibilities.

To understand the effects of the tides on hot Jupiters and their orbital evolution, however,
we need to know the amount of dissipations caused by the tides in their interiors.
The amount of tidal dissipations is sometimes parametrized by introducing the tidal quality factor $Q$, 
which is however difficult to theoretically determine because of lack of
our knowledge of the thermal properties of the matter in the planet interior
and of how convective fluids respond to the tidal forcing.
Tidal responses of the planets and stars may depend on their structure, dissipation mechanisms
in the interior, their rotation rates,
the forcing frequency, and so on. 
There are cases in which the $Q$ value is observationally determined. 
For example, the $Q$ value for Jupiter is estimated to be order of $\sim10^5$ from observations of the satellite Io orbiting the planet (e.g., Goldreich \& Soter 1966).

Giant gas planets are in general composed of a geometrically thin radiative envelope and a convective core with or without a central solid region (e.g., Guillot et al 2004; Guillot 2005).
The responses of convective fluids in the core to the tidal forces are
difficult to model theoretically.
Fluid motions in the convective core may be affected by rotation rates of the planets (e.g., Stevenson 1979).
The tidal responses of the convective fluids may also depend on the forcing frequency 
(e.g., Zahn 1966, 1989; Goldreich \& Keeley 1977).
A way to model tidal responses of the convective fluids may be to treat the fluids in the convective core as 
viscous fluids with effective viscosities due to turbulent convection based on
the mixing length theory.
For example, to compute tidal responses of non-rotating solar type stars with the convective envelope,
Terquem et al (1998) solved linearized equations of motion with a viscous term for the envelope where
the turbulent viscosity coefficient is assumed to depend on the ratio
of the convective timescale to the tidal forcing period (see, e.g., Xion, Cheng \& Deng 1997).
Note that they took account of the effects of radiative dissipation on the responses.
For uniformly rotating solar type stars, Savonije \& Witte (2002) (see also Savonije \& Papaloizou 1997) computed 
non-adiabatic tidal responses taking account of viscous dissipation in the envelope, where they employed
the traditional approximation to describe the oscillations of rotating stars (e.g., Lee \& Saio 1987, 1997).
For Jovian planets, Ogilvie \& Lin (2004) computed inertial modes of rotating convective planets 
with a solid core treating the convective fluid as a viscous fluid, where the convective interior 
is assumed to be nearly isentropic.
They investigated the properties of tidal responses taking account of viscous dissipation caused by the inertial modes and of energy leakages by traveling low frequency $g$-modes, where the Ekman number was assumed a constant parameter for the viscosity coefficient.
Focusing on inertial modes, Wu (2005) also studied the properties of tidal dissipations in uniformly rotating giant planet
and gave an estimation to the tidal $Q$ factor, 
where the viscosity coefficient similar to that used by Terquem et al (1998) was employed.
Although no solid core was assumed at the centre,
Wu (2005) derived similar conclusions to those by Ogilvie \& Lin (2004).
To investigate dynamical tides in fully convective planets, 
Papaloizou \& Ivanov (2010) carried out numerical simulations of close encounters between 
a rotating Jovian planet and 
a central star, assuming the convective fluids as viscous fluids with a constant kinetic viscosity coefficient.
Papaloizou \& Ivanov (2010) suggested that the total energy exchanged during the encounter does not strongly depend on the existence of a solid core in the convective interior.

Thermal tides in hot Jupiters were discussed by Arras \& Socrates (2010),
Auclair-Desrotour \& Leconte (2018), and Lee \& Murakami (2019) as a mechanism that keeps
the rotation of hot Jupiters asynchronous to the orbital motion so that tidal heatings in the interior can be
operative to inflate the radii of the planets.
Arras \& Socrates (2010), for non-rotating hot Jupiters, computed tidal responses to 
semi-diurnal insolation of the hosting star.
Auclair-Desrotour \& Leconte (2018) computed for uniformly rotating hot Jupiters
non-adiabatic responses using the traditional approximation to describe the responses in rotating planets.
They showed that depending on the tidal forcing periods 
the tides can produce the torques which keep the spin of the planet asynchronous to the orbital motion.
Lee \& Murakami (2019) revisited the problem of thermal tides in uniformly rotating Jovian planets,
without using the traditional approximation, and they
computed non-adiabatic tidal responses, assuming that the convective core has 
an almost isentropic structure so that the core supports propagation of inertial modes.
Lee \& Murakami (2019) confirmed most of the results obtained by Auclair-Desrotour \& Leconte (2018), but
they also showed that thermal tides can excite low order inertial modes in the convective core and $r$-modes
in the envelope, which could modify the behavior of the tidal torques.

In this paper, 
treating the convective fluids in the core of hot Jupiters as viscous fluids,  
we compute small amplitude gravitational and thermal tides in uniformly rotating Jovian planets
for simple hot Jupiter models composed of a thin radiative envelope and a nearly isentropic convective core.
This paper is an extension of Lee \& Murakami (2019) to the case in which the core convective fluids
are treated as viscous fluids.
For the forcing terms, we consider both gravitational tidal potential and semi-diurnal insolation
due to the central star.
As dissipation mechanisms responsible for tidal torques in the planet interior, 
we consider radiative dissipation in the envelope and viscous dissipation in the convective core where
we assume the Ekman number as a constant parameter to derive viscous coefficient
necessary to compute viscous dissipations in the core (Ogilvie \& Lin 2004).
\S 2 is for numerical method used in this paper, and \S 3 gives numerical results.
Conclusions will be given in \S 4.
The Appendices describe the derivation of the oscillation equations solved for the convective core
and the expression used to calculate viscous energy dissipation rates.

\section{Method of Solution}

We calculate small amplitude gravitational and thermal tides of uniformly rotating hot Jupiters, composed of
a thin radiative envelope and a convective core (see, e.g.,  Arras \& Socrates 2010;
Auclair-Desrotour \& Leconte 2018; Lee \& Murakami 2019), taking
account of viscous dissipations in the core and
radiative dissipations in the envelope.
We linearize the basic equations for the fluid dynamics to describe small amplitude tidal responses.

\subsection{Equations for the Viscous Core}

\subsubsection{Basic Equations for viscous fluids}

Regarding convective fluids in the core as viscous fluids with turbulent viscosity,
we use linearized Navier-Stokes equation.
The basic equations governing viscous fluids
in a uniformly rotating planet driven by the tidal potential $\Phi_T$ due to the host star
are given in the co-rotating frame of the planet by
 (e.g., Landau \& Lifshitz 1987)
\be
\rho\left(\frac{d\pmb{v}}{d t}+2\pmb{\Omega}\times\pmb{v}\right)=-\nabla p-\rho\nabla\Phi-\rho\nabla\Phi_T+\nabla\cdot\pmbmt{\sigma},
\label{eq:fluiddyn}
\ee
\be
\frac{\partial\rho}{\partial t}+\nabla\cdot\left(\rho\pmb{v}\right)=0,
\label{eq:continuity}
\ee
\be
\rho T \frac{ds}{dt}=\rho\epsilon-\nabla\cdot\pmb{F}+\sum_{i,j}\sigma_{ij}{\partial v_i\over\partial x_j},
\ee
where $\pmb{v}$ is the velocity vector of the fluid, $\pmbmt{\sigma}=(\sigma_{ij})$ is the symmetric viscous stress tensor,
$p$ is the pressure,
$\rho$ is the mass density, 
$T$ is the temperature, $s$ is the specific entropy, $\pmb{F}$ is the energy flux vector, 
$\Phi$ is the gravitational
potential, and
$\epsilon$ is the energy generation rate per gram.
We ignore rotational deformation of the planet for simplicity.
The angular velocity of rotation $\pmb{\Omega}$ is assumed constant and is along the $z$-axis.
Using the shear and bulk viscosity coefficients $\eta$ and $\zeta$, we may write the stress tensor $\sigma_{ij}$
as (e.g., Landau \& Lifshitz 1987)
\be
\sigma_{ij}=\zeta\left(\nabla\cdot\pmb{v}\right)\delta_{ij}+2\eta\left(e_{ij}-{1\over 3}\delta_{ij}\nabla\cdot\pmb{v}\right), 
\ee
where
\be
e_{ij}={1\over 2}\left({\partial v_j\over\partial x_i}+{\partial v_i\over\partial x_j}\right),
\ee
and $\delta_{ij}$ is the Kronecker symbol.
The components of the stress tensor in spherical polar coordinates $(r,\theta,\phi)$, whose origin
is at the centre of the planet, are given by
\be
\sigma_{rr}=\left(\zeta-{2\over 3}\eta\right)\nabla\cdot\pmb{v}+2\eta{\partial v_r\over\partial r}, 
\ee
\be
\sigma_{\theta\theta}=\left(\zeta-{2\over 3}\eta\right)\nabla\cdot\pmb{v}+2\eta\left({1\over r}{\partial v_\theta\over\partial\theta}+{v_r\over r}\right), \ee
\be
\sigma_{\phi\phi}=\left(\zeta-{2\over 3}\eta\right)\nabla\cdot\pmb{v}+2\eta\left({1\over r\sin\theta}{\partial v_\phi\over\partial\phi}
+{v_r\over r}+{v_\theta\cot\theta\over r}\right), 
\ee
\be
\sigma_{r\theta}=\eta\left({1\over r}{\partial v_r\over\partial\theta}+r{\partial\over\partial r}{v_\theta\over  r}\right), 
\ee
\be
\sigma_{r\phi}=\eta\left({1\over r\sin\theta}{\partial v_r\over \partial\phi}+r{\partial\over\partial r}{v_\phi\over r}\right),
\ee
\be
\sigma_{\theta\phi}=\eta\left({1\over r\sin\theta}{\partial v_\theta\over\partial\phi}+{\sin\theta\over r}{\partial\over\partial\theta}
{v_\phi\over\sin\theta}\right).
\ee

\subsubsection{Perturbed Basic Equations}

Regarding the tidal potential $\Phi_T$ as a small amplitude perturbation, 
we obtain, for small amplitude tidal responses, a set of perturbed basic equations, which are
\be
{\partial\pmb{v}'\over\partial t}+2\pmb{\Omega}\times\pmb{v}'=-{1\over\rho}\nabla p'-{\rho'\over\rho}\nabla \Phi-\nabla\Phi_T
+{1\over\rho}\nabla\cdot\pmb{\sigma}',
\label{eq:linearizedeom}
\ee
\be
{\partial\rho'\over\partial t}+\nabla\cdot\left(\rho\pmb{v}'\right)=0,
\label{eq:linearizedcont}
\ee
\be
\rho T {\partial\over \partial t}\delta s=(\rho\epsilon)'-\nabla\cdot\pmb{F}',
\label{eq:linearizedentropy}
\ee
where the prime $(')$ and $\delta$ indicate the Eulerian and Lagrangian perturbations, respectively, and
we have employed the Cowling approximation, in which the Eulerian perturbation $\Phi'$ of the gravitational potential 
and the perturbed Poisson equation are ignored.
Note that for uniform rotation the viscous term $(\sigma_{ij}\partial v_i/\partial x_j)^\prime$ is the second order in the perturbation amplitudes and does not appear in the first order perturbed entropy equation.
Since we ignore rotational deformation of the equilibrium structure, the planets are assumed to be
spherical symmetric and the hydrostatic equation is simply given by $dp/dr=-\rho d\Phi/dr$.

We treat the tidal potential and insolation due to the host star as small amplitude perturbations.
Assuming the rotation axis of the planet is perpendicular to the orbital plane, 
the tidal potential $\Phi_T(\pmb{r})=-{GM_*/ |\pmb{r}-\pmb{a}_*(t)|}$ in the planet due to the host star with the mass $M_*$ may be expanded as (e.g., Press \& Teukolsky 1977)
\be
\Phi_T(\pmb{r})
=-GM_*\sum_{lm}W_{lm}{r^l\over a_*(t)^{l+1}}e^{-\rmi mE}Y_l^m(\theta,\phi)\equiv\sum_{lm}\Phi^{(lm)}_T,
\label{eq:phit0}
\ee
where $G$ is the gravitational constant, $Y_l^m$ is the spherical harmonic function,
$\pmb{a}_*(t)$ is the position vector to the host star, $a_*(t)=|\pmb{a}_*(t)|$, $E$ is the true anomaly, and
\be
W_{lm}=(-1)^{(l+m)/2}{\displaystyle\left[{4\pi\over 2l+1}(l-m)!(l+m)!\right]^{1/2}\over\displaystyle
\left[2^l\left({l-m\over 2}\right)!\left({l+m\over 2}\right)!\right]}=W_{l,-m},
\ee
which has non-zero values for even integers of $l-m$.
Note that the tidal potential $\Phi_T$ does not contain the dipole terms associated with $l=1$.
Assuming that the eccentricity $e$ of the orbit is small so that $E\approx \Omega_{\rm orb}t$, and taking only the dominant tidal component with $l=-m=2$, the tidal potential $\Phi_T$ in an inertial frame may be given by
\be
\Phi_T^{(2,-2)}=-\sqrt{3\pi\over 10}{GM_*\over a_*^3}r^2Y_2^{-2}(\theta,\phi)e^{2\rmi \Omega_{\rm orb}t}
\equiv \Phi_T^{(2)}Y_2^{-2}(\theta,\phi)e^{2\rmi \Omega_{\rm orb}t},
\label{eq:phit}
\ee
where $W_{2,-2}=\sqrt{3\pi/ 10}$ and $\Omega_{\rm orb}=\sqrt{G(M_*+M)/a_*^3}$ with $M$ being the mass of the planet is the mean angular velocity of the orbital motion.
When the planet is uniformly rotating at the angular velocity $\Omega$, the forcing frequency $\omega$ in the co-rotating frame of the planet
may be given by  
$\omega=2\Omega_{\rm orb}+m\Omega=2(\Omega_{\rm orb}-\Omega)$ for $m=-2$.

Thermal tides are caused by insolation by the host star, which 
produces the day and night sides on the
planet and  
is given by 
(e.g., Auclair-Desrotour \& Leconte 2018) 
\be
\epsilon'=\left\{\begin{array}{ll}
J_*(r,\theta,\phi,t)=\kappa_*F_*e^{-p(r)/p_*}\cos\phi_* & {\rm for} \quad 0\le\phi_*\le\pi/2\\
J_*(r,\theta,\phi,t)=0 & {\rm for} \quad \pi/2\le\phi_*\le\pi,
\end{array}\right.
\label{eq:epsilonprime0}
\ee
where $\phi_*$ is the zenith angle of the host star as observed from the planet and is
given by $\cos\phi_*=\sin\theta\cos(\phi-E)$, and
$F_*=\sigma_{\rm SB}T_*^4\left({R_*/ r_*}\right)^2$, $\kappa_*={g_0/ p_*}$, $g_0={GM/ R^2}$,
and $T_*$ and $R_*$ are respectively the surface temperature and radius of the host star,  $p_*$ and 
$\kappa_*$ are the pressure and the opacity at the base of the heated layer in the planet,
and $r_*$ is the distance between the planet and the host star, set equal to the semi-major axis $a_*$ of the orbit.
In the co-rotating frame of the planet, we may expand $J_*$ in terms of spherical harmonic function as
\be
J_*(r,\theta,\phi,t)=\sum_{n,l,m}J^{(nlm)}_*(r)Y_l^m(\theta,\phi)e^{\rmi\omega_{nm} t},
\label{eq:epsilonexpnasion}
\ee
where
$\omega_{nm}=n\Omega_{\rm orb}+m\Omega$ and $n\Omega_{\rm orb}$ represents the forcing frequency in an
inertial frame.
If we take only the component with $n=-m=2$, for which $\omega_{2,-2}=2(\Omega_{\rm orb}-\Omega)$, 
we obtain
\be
\epsilon'=J_*^{(2,2,-2)}Y_2^{-2}(\theta,\phi)e^{\rmi\omega t}
={1\over 16}\sqrt{15\pi\over 2}\kappa_*F_*e^{-p(r)/p_*}Y_2^{-2}(\theta,\phi)e^{\rmi\omega t},
\ee
where $\omega=\omega_{2,-2}$.

The time dependence of the responses to the tidal forcing with the forcing frequency $\omega$ in the co-rotating frame of the planet
may be given by the factor $e^{\rmi\omega t}$.
For uniformly rotating planets, we have
\be
\pmb{v}'=\rmi\omega\pmb{\xi},
\ee
where $\pmb{\xi}$ is the displacement vector given by $\pmb{\xi}=\xi_r\pmb{e}_r+\xi_\theta\pmb{e}_\theta+\xi_\phi\pmb{e}_\phi$ in the spherical polar coordinates, 
and $\pmb{e}_r$, $\pmb{e}_\theta$, and $\pmb{e}_\phi$ are the orthonormal vectors in the $r$, $\theta$, and $\phi$
directions, respectively.
The perturbed equation of motion is given by
\be
-\omega^2\pmb{\xi}+2\rmi\omega\pmb{\Omega}\times\pmb{\xi}=-{1\over\rho}\nabla p'-{\rho'\over\rho}\nabla \Phi
-\nabla\Phi_T
+{1\over\rho}\nabla\cdot\pmb{\sigma}',
\label{eq:linearizedeom}
\ee
and linearization of the continuity equation (\ref{eq:linearizedcont}) and the entropy equation (\ref{eq:linearizedentropy}) leads respectively to 
\be
\rho'+\nabla\cdot\left(\rho\pmb{\xi}\right)=0,
\ee
and 
\be
\rmi\omega\rho T\delta s=(\rho\epsilon)'-\nabla\cdot\pmb{F}'.
\label{eq:dsdt}
\ee
The equation of state $p=p(\rho,s)$ is perturbed to be
\be
{\rho'\over\rho}=-rA{\xi_r\over r}+{1\over\Gamma_1}{ p'\over p}-\alpha_T{\delta s\over c_p},
\label{eq:eos}
\ee
where $c_p$ is the specific heat at constant pressure, 
$\Gamma_1=\left({\partial\ln p/\partial\ln\rho}\right)_{\rm ad}$, 
$\alpha_T=-\left({\partial\ln\rho/\partial\ln T}\right)_p$,
and
\be
rA={d\ln\rho\over d\ln r}-{1\over\Gamma_1}{d\ln p\over d\ln r}.
\ee

For the perturbed entropy equation (\ref{eq:dsdt}), we employ the approximation used by
Auclair-Desrotour \& Leconte (2018), who assumed Newtonian cooling in the envelope.
The second term on the right-hand-side of equation (\ref{eq:dsdt}) may be approximately given by
\be
\nabla\cdot\pmb{F}'
\sim { \omega_D\rho c_p T}{T'\over T}
=\omega_D\rho c_p T\left({\delta s\over c_p}+\nabla_{\rm ad}{\delta p\over p}+\nabla V{\xi_r\over r}\right),
\ee
where 
$V=-{d\ln p/ d\ln r}$, $\nabla={d\ln T/ d\ln p}$, $\nabla_{\rm ad}=\left({d\ln T/ d\ln p}\right)_{\rm ad}$,
and
\be
\omega_D={4\pi\over\tau_*}\left[\left({p\over p_*}\right)^{1/2}+\left({p\over p_*}\right)^2\right]^{-1},
\label{eq:omegad}
\ee
where $\tau_*$ is the timescale parameter to specify the efficiency
of radiative cooling in the envelope, and we use $p_*=10^6{\rm dyn/cm^2}$ (see Auclair-Desrotour \& Leconte 2018; Iro et al 2005).
The entropy perturbation $\delta s$ is then given by
\be
{\delta s\over c_p}={1\over\rmi\omega+\omega_D}{(\rho\epsilon)'\over \rho Tc_p}-{\omega_D\over\rmi\omega+\omega_D}\left[\nabla_{\rm ad}
V\left({p'_l\over \rho gr}-{\xi_r\over r}\right)+\nabla V{\xi_r\over r}\right].
\ee
The density perturbation is now rewritten as
\be
{\rho'\over\rho}
=\left(-rA-\beta_1\right){\xi_r\over r}+\left({V\over\Gamma_1}+\beta_2\right){p'\over\rho gr}-{\alpha_T\over\rmi\omega+\omega_D}
{\epsilon'\over Tc_p},
\ee
where
\be
\beta_1={\alpha_T(\nabla_{\rm ad}-\nabla)V\over\rmi\omega/\omega_D+1}, \quad
\beta_2={\alpha_T\nabla_{\rm ad}V\over\rmi\omega/\omega_D+1},
\ee
and we have used $\rho'\epsilon=0$ since no insolation is assumed in equilibrium state.

\subsubsection{Oscillation Equations for Viscous Fluids}

Since separation of variables is not possible for the perturbations in rotating stars,
assuming that the equilibrium state is axisymmetric about the rotation axis, we use 
finite series expansions in terms of spherical harmonic functions $Y_l^m(\theta,\phi)$
with different $l$s for a given $m$.
The three components of the displacement vector $\pmb{\xi}(\pmb{x},t)$ are given by
\be
{\xi_r}=r\sum_{j=1}^{j_{\rm max}}S_{l_j}(r)Y_{l_j}^m(\theta,\phi)e^{\rmi\omega t},
\label{eq:xiexp_r}
\ee
\be
{\xi_\theta}=r\sum_{j=1}^{j_{\rm max}}\left[H_{l_j}(r)\frac{\partial}{\partial\theta}
Y_{l_j}^m(\theta,\phi)+T_{l'_j}(r)\frac{1}{\sin\theta}\frac{\partial}{\partial\phi}Y_{l'_j}^m(\theta,\phi)
\right]e^{\rmi\omega t},
\label{eq:xiexp_theta}
\ee
\be
{\xi_\phi}=r\sum_{j=1}^{j_{\rm max}}\left[H_{l_j}(r)\frac{1}{\sin\theta}\frac{\partial}{\partial\phi}
Y_{l_j}^m(\theta,\phi) - T_{l'_j}(r)\frac{\partial}{\partial\theta}Y_{l'_j}^m(\theta,\phi)
\right]e^{\rmi\omega t},
\label{eq:xiexp_phi}
\ee
and the Eulerian pressure perturbation,
$p^\prime(\pmb{x},t)$, is given by
\be
p^\prime=\sum_{j=1}^{j_{\rm max}}p^\prime_{l_j} (r)Y_{l_j}^m\left(\theta,\phi\right)e^{\rmi\omega t},
\label{eq:pexp}
\ee
where $l_j=2(j-1)+|m|$ and $l'_j=l_j+1$ for even modes, and $l_j=2j-1+|m|$ and
$l'_j=l_j-1$ for odd modes for $j=1~,2~,3\cdots,~j_{\rm max}$ (e.g., Lee \& Saio 1986).
Since the angular dependence of the tidal forces is
given by $Y_l^m(\theta,\phi)$
with $l=-m=2$, the tidally forced oscillations have even parity such that $l_j=2j$ and $l'_j=l_j+1$
for the expansion coefficients.

Substituting the expansions as given by equations (\ref{eq:xiexp_r}) to (\ref{eq:pexp}) into the perturbed basic
equations, we obtain
a set of linear ordinary differential equations for the expansion coefficients with the forcing terms
due to the tidal potential and insolation from the host star.
Defining the dependent variables $\pmb{z}_j$ for $j=1$ to $6$ as
\be
\pmb{z}_1=\left(S_l\right), \quad \pmb{z}_2=\left(z_{2,l}\right), \quad \pmb{z}_3=\left(H_l\right), \quad 
\pmb{z}_4=\left(\alpha_1\left(r{\partial\over\partial r}H_l+S_l\right)\right), \quad \pmb{z}_5=\left(iT_{l'}\right), \quad
\pmb{z}_6=\alpha_1r{\partial\over\partial r}\pmb{y}_5, 
\ee
where $z_{2,l}$ is defined by equation (A7), and introducing
\be
\pmb{\psi}=(\Phi_{T}^{(l)}), \quad \pmb{Z}_2=\pmb{z}_2+{\pmb{\psi}\over gr},
\quad \pmb{j}_*=\left({J_*^{(l)}\over Tc_p}\right),
\ee
where non-zero $\Phi_T^{(l)}$ and $J_*^{(l)}\equiv J_*^{(2,l,m)}$ occur only for $l=2$, 
we obtain the oscillation equations for rotating viscous planets:
\be
r{\partial\pmb{z}_1\over \partial r}=\left(1+\rmi{\alpha_2+4\alpha_1/3\over\hat\Gamma_1}\right)^{-1}\left[
\left({V\over\hat\Gamma_1}-3-3\rmi{\alpha_2\over\hat\Gamma_1}\right)\pmb{z}_1-{V\over\hat\Gamma_1}\pmb{Z}_{2}
+\left(1+\rmi{\alpha_2-2\alpha_1/3\over\hat\Gamma_1}\right)\pmbmt{\Lambda}_0\pmb{z}_3
+{V\over\hat\Gamma_1}{\pmb{\psi}\over gr}+{\alpha_T\over\rmi\omega+\omega_D}\pmb{j}_*\right],
\label{eq:y1}
\ee
\begin{eqnarray}
r{\partial\pmb{Z}_2\over \partial r}&=&\left(c_1\bar\omega^2+rA+\beta_1\right)\pmb{z}_1
+\left(1-U-rA-\beta_2\right)\pmb{Z}_2
-2mc_1\bar\omega\bar\Omega \pmb{z}_3-2c_1\bar\omega\bar\Omega \pmbmt{C}_0\pmb{z}_5
+\left(rA+\beta_2\right){\pmb{\psi}\over gr}+{\alpha_T\over\rmi\omega+\omega_D}\pmb{j}_*
\nonumber\\
&&-\rmi\left[\left({\alpha_2+4\alpha_1/3\over\hat\Gamma_1}-4{\alpha_1\over V}\right)r{\partial\pmb{z}_1\over\partial r}
+{3\alpha_2\over\hat\Gamma_1}\pmb{z}_1-\left({\alpha_2-2\alpha_1/3\over\hat\Gamma_1}+2{\alpha_1\over V}\right)\pmbmt{\Lambda}_0\pmb{z}_3
+{1\over V}\pmbmt{\Lambda}_0\pmb{z}_4\right],
\label{eq:y2}
\end{eqnarray}
\be
r{\partial\pmb{z}_3\over\partial r}=-\pmb{z}_1+{1\over\alpha_1}\pmb{z}_4,
\label{eq:y3}
\ee
\be
r{\partial\pmb{z}_4\over\partial r}=-2\alpha_1\left(\pmbmt{I}-\pmbmt{\Lambda}_0\right)\pmb{z}_3+\left(V-3\right)\pmb{z}_4+2\alpha_1r{\partial\pmb{z}_1\over\partial r}+\rmi Vc_1\bar\omega^2\left(\pmbmt{L}_0\pmb{z}_3-\pmbmt{M}_1\pmb{z}_5-mq\pmbmt{\Lambda}_0^{-1}\pmb{z}_1-{\pmb{Z}_2\over c_1\bar\omega^2}\right) +\rmi V{\pmb{\psi}\over gr},
\label{eq:y4}
\ee
\be
r{\partial\pmb{z}_5\over\partial r}={1\over\alpha_1}\pmb{z}_6,
\label{eq:y5}
\ee
\be
r{\partial\pmb{z}_6\over\partial r}=-\alpha_1\left(2\pmbmt{I}-\pmbmt{\Lambda}_1\right)\pmb{z}_5+\left(V-3\right)\pmb{z}_6
{+}\rmi Vc_1\bar\omega^2\left(\pmbmt{L}_1\pmb{z}_5-\pmbmt{M}_0\pmb{z}_3+q\pmbmt{K}\pmb{z}_1\right),
\label{eq:y6}
\ee
where $c_1={(r/R)^3/ (M_r/M)}$, $U={d\ln M_r/ d\ln r}$, and 
\be
\bar\omega={\omega\over\sigma_0}, \quad 
\bar\Omega={\Omega\over\sigma_0}, \quad \sigma_0=\sqrt{GM\over R^3}, \quad q={2\Omega\over \omega}, 
\ee 
\be
\alpha_1={\omega\eta\over p}, \quad \alpha_2={\omega\zeta\over p},
\quad {1\over\hat\Gamma_1}={1\over\Gamma_1}+{\beta_2\over V},
\ee
and $\pmbmt{I}$ is the unit matrix and 
the non-zero elements of the matrices $\pmbmt{\Lambda}_0$, $\pmbmt{\Lambda}_1$, $\pmbmt{L}_0$, $\pmbmt{L}_1$,
$\pmbmt{M}_0$, $\pmbmt{M}_1$, $\pmbmt{K}$, and $\pmbmt{C}_0$ for even modes with $l_j=2(j-1)+|m|$ and $l_j'=l_j+1$ for $j=1,~2,~,\cdots,~j_{\rm max}$ are given by
\be
\left(\pmbmt{\Lambda}_0\right)_{j,j}=l_j(l_j+1),\quad \left(\pmbmt{\Lambda}_1\right)_{j,j}=l'_j(l'_j+1),\quad
\left(\pmbmt{L}_0\right)_{j,j}=1-{mq\over l_j(l_j+1)}, \quad \left(\pmbmt{L}_1\right)_{j,j}=1-{mq\over l'_j(l'_j+1)},
\ee
\be
\left(\pmbmt{M}_0\right)_{j,j}=q{l_j\over l_j+1}J^m_{l_j+1}, \quad \left(\pmbmt{M}_0\right)_{j,j+1}=q{l_j+3\over l_j+2}J^m_{l_j+2}, \quad
\left(\pmbmt{M}_1\right)_{j,j}=q{l_j+2\over l_j+1}J^m_{l_j+1}, \quad \left(\pmbmt{M}_1\right)_{j+1,j}=q{l_j+1\over l_j+2}J^m_{l_j+2},
\ee
\be
\left(\pmbmt{K}\right)_{j,j}={J^m_{l_j+1}\over l_j+1}, \quad \left(\pmbmt{K}\right)_{j,j+1}=-{J^m_{l_j+2}\over l_j+2},
\quad \left(\pmbmt{C}_0\right)_{j,j}=-(l_j+2)J_{l_j+1}^m, \quad \left(\pmbmt{C}_0\right)_{j+1,j}=(l_j+1)J_{l_j+2}^m,
\ee
where
$
J^m_l=\sqrt{(l^2-m^2)/ (4l^2-1)}
$
for $l\ge|m|$ and $J^m_l=0$ otherwise (e.g., Lee \& Saio 1990).

In this paper, we ignore the bulk viscosity and set
$\zeta=0$ and hence $\alpha_2=0$ for simplicity.
For rotating planets, we may define the Ekman number $\rm Ek$ as
\be
{\rm Ek}={\nu\over 2|\Omega| R^2},
\ee
where $\nu=\eta/\rho$ is the kinematic viscosity, $R$ is the radius of the planet.
Assuming that the Ekman number $\rm Ek$ is a constant parameter,
we obtain
\be
\nu=2|\Omega| R^2{\rm Ek}, \quad \alpha_1=2c_1{ \bar\omega|\bar\Omega|}{V\over x^2}{\rm Ek},
\ee
where $x=r/R$.

{Using the numbers given in Table 3.4 by Guillot et al (2004) for physical quantities in Jupiter, we find
the microscopic viscosity $\nu\sim 10^{-2}{\rm cm^2/s}$, which leads to an Ekman number ${\rm Ek}\sim 10^{-17}$.
If we use turbulent viscosity in the convective core $\nu_{\rm turb}\sim l_mv_{\rm turb}$,
instead of the microscopic viscosity $\nu$, from the same table we find the mean free path $l_m\sim H_p\sim3\times10^9{\rm cm}$ and
the turbulent velocity $v_{\rm turb}\sim v_{\rm conv}\sim3~{\rm cm/s}$, where $H_p$ is the pressure scale height and 
$v_{\rm conv}$ the turbulent velocity of convective fluid elements, and hence ${\rm Ek}\sim 10^{-7}$, which
we use in this paper as a typical value of the Ekman number in the convective core of hot Jupiters.}

\subsection{Equations for the radiative envelope}

The oscillation equations solved in the radiative envelope are the same as
those given by Lee \& Murakami (2019) and they are
\be
r{d\pmb{y}_1\over dr}
=\left[\left({V\over\hat\Gamma_1}-3\right)\pmbmt{1}+q\pmbmt{W}\pmbmt{O}\right]\pmb{y}_1
+\left({\pmbmt{W}\over c_1\bar\omega^2}-{V\over\hat\Gamma_1}\pmbmt{1}\right)\pmb{Y}_2
+{\alpha_T\over\rmi\omega+\omega_D}\pmb{j}_*+{V\over\hat\Gamma_1}{\pmb{\psi}\over gr},
\label{eq:dy1}
\ee
\be
r{d\pmb{Y}_2\over dr}
=\left[\left(c_1\bar\omega^2+rA+\beta_1\right)\pmbmt{1}-4c_1\bar\Omega^2\pmbmt{G}\right]\pmb{y}_1
+\left[\left(1-U-rA-\beta_2\right)\pmbmt{1}-q\pmbmt{O}^T\pmbmt{W}\right]\pmb{Y}_2
+{\alpha_T\over\rmi\omega+\omega_D}\pmb{j}_*+\left(rA+\beta_2\right){\pmb{\psi}\over gr},
\label{eq:dy2}
\ee
where 
\be
\pmb{y}_1=(S_l), \quad \pmb{y}_2=\left({p'_l\over \rho gr}\right), \quad \pmb{Y}_2=\pmb{y}_2+{\pmb{\psi}\over gr},
\ee
and 
\be
\pmbmt{W}=\pmbmt{\Lambda}_0(\pmbmt{L}_0-\pmbmt{M}_1\pmbmt{L}_1^{-1}\pmbmt{M}_0)^{-1}, \quad 
\pmbmt{O}=m\pmbmt{\Lambda}_0^{-1}-\pmbmt{M}_1\pmbmt{L}_1^{-1}\pmbmt{K}, \quad 
\pmbmt{G}=\pmbmt{O}^T\pmbmt{W}\pmbmt{O}-\pmbmt{C}_0\pmbmt{L}_1^{-1}\pmbmt{K},
\ee 
and $\pmbmt{O}^T$ is the transposed matrix of $\pmbmt{O}$.

{To solve the sets of linear ordinary differential equations derived in \S 2.1 and 2.2, we use 
a Henyey type relaxation method,
the detail of which may be found in Unno et al (1989).}

\subsection{Boundary Conditions and Jump Conditions}

The inner boundary conditions applied at the centre of the planet are given by
\be
v_r'=0, \quad v_\theta'=0, \quad v_\phi'=0,
\label{eq:ibdc}
\ee
for which we use $\pmb{z}_1=\pmb{z}_3=\pmb{z}_5=0$.
The outer boundary condition at the surface of the planet is given by $\delta p=0$, for which we use $\pmb{y}_2-\pmb{y}_1=0$.

At the interface between the inviscid envelope and the viscous core, we assume the horizontal components 
of the stress tensor vanish so that
\be
\sigma'_{r\theta}=\rmi\omega\eta\left[\sum_l\left(d{dH_l\over dr}+S_l\right){\partial Y_l^m\over
\partial\theta}+\sum_{l'}r{dT_{l'}\over dr}{1\over \sin\theta}{\partial Y_{l'}^m\over\partial\phi}\right]=0,
\ee
\be
\sigma'_{r\phi}=\rmi\omega\eta\left[\sum_l\left(r{dH_l\over dr}+S_l\right){1\over\sin\theta}{\partial Y_l^m\over\partial\phi}-\sum_{l'}r{dT_{l'}\over dr}{\partial Y_{l'}^m\over\partial\theta}\right]=0,
\ee
which leads to
\be
\pmb{z}_4=0, \quad \pmb{z}_6=0.
\ee
We also use at the interface
\be
\pmb{z}_1=\pmb{y}_1, \quad \pmb{z}_2=\pmb{y}_2.
\ee

\section{Numerical Results}

\subsection{Equilibrium Models}

In this paper, we use simple equilibrium models of hot Jupiters computed with the equation of state given by
(see, Arras \& Socrates 2010; Auclair-Desrotour \& Leconte 2018)
\be
\rho(p)=e^{-p/p_b}{p\over a^2}+\left(1-e^{-p/p_b}\right)\sqrt{p\over K_c},
\ee
where
\be
K_c=GR_J^2, \quad a^2=\sqrt{p_bK_c},
\ee 
and $p_b$ is a pressure parameter used to define the core-envelope boundary
and $a$ is the isothermal sound velocity and $R_J$ is the radius of Jupiter.
The models consist of a convective core and a radiative envelope, which is nearly isothermal.
Assuming $\Gamma_1\equiv \left({\partial\ln p/\partial\ln\rho}\right)_{ad}=2$, we obtain 
a nearly isentropic convective core (e.g., Stevenson \& Salpeter 1977a,b).
In this paper, we use $p_b=100{\rm bar}=10^8{\rm dyn/cm^2}$ and set the outer boundary $R_e$ at
$p=0.01{\rm dyn/cm^2}$ and define the planet's radius $R=R_e/1.01$.
We use $M=0.7M_J$ and the radius $R=9.31\times10^9$cm.
For this model, the core-envelope boundary is at $x=r/R=0.957$ corresponding to $p=1.89\times10^8 ({\rm dyn/cm^2})$.
For numerical computations, we assume $M_*=M_\odot$, $R_*=R_\odot$, and $T_*=5.8\times10^3$K for the host star, and the distance between the planet and the star is assumed to be $r_*=a_*=0.05$A.U..
Note that for the parameter values we use, we have $|\bar\Omega_{\rm orb}|=0.0537$.

\subsection{Core Inertial Modes}

It is useful to compute inertial modes in the viscous core as eigen-modes.
Inertial modes are rotationally induced oscillation modes, for which 
the Coriolis force is the restoring force.
There exist prograde and retrograde inertial modes and in our convention
prograde (retrograde) modes have $m\omega<0$ ($m\omega>0$). 
Note that so called $r$-modes form a subclass of inertial modes and appear only on the 
retrograde side.
The frequency $\omega$ of inertial modes and $r$-modes is proportional to the rotation speed $\Omega$ and
the ratio $\omega/\Omega$ of an inertial mode tends to a constant value in the limit of $\Omega\rightarrow 0$
(e.g., Yoshida \& Lee 2000; see also Lockitch \& Friedman 1999).
Note that for a given rotation speed $\Omega$, the frequency $\omega$ of inertial modes is limited to the inertial range given by
$|\omega|\le 2\Omega$.

To compute eigen-modes by solving the set of linear differential equations derived in \S 2, 
we set $\pmb{j}_*=0$ and $\pmb{\psi}=0$ and introduce a normalization
condition for the amplitudes.
It is important to note that the frequency $\omega$ of non-radial modes becomes complex such that
$\omega=\omega_{\rm R}+\rmi\omega_{\rm I}$ because of radiative
dissipations in the envelope and viscous dissipations in the core.
When the time dependence of the perturbations is given by the factor $e^{\rmi \omega t}$,
the eigenmodes with $\omega_{\rm I}={\rm Im}(\omega)>0$ are stable.

{For modal analyses made in this paper,
we use a planet model that has about 1000 mech points and the expansion length $j_{\rm max}=10$ for the perturbations.
We have computed free inertial modes and tidal responses also for $j_{\rm max}=15$ for $\tau_*=1$ day and Ek$=10^{-7}$ and
found no essential differences from those for $j_{\rm max}=10$.}

In Table 1 we tabulate the complex eigenfrequency $\bar\omega=\bar\omega_{\rm R}+{\rm i}\bar\omega_{\rm I}$
of $m=-2$ inertial modes of even parity for ${\rm Ek}=10^{-9}$, $10^{-7}$, and $10^{-5}$, where
we assume $\tau_*=1$ day and $\bar\Omega=0.1$. Here,
the inertial modes are labeled as $i_{l_0-|m|}$ 
(see, e.g., Yoshida \& Lee (2000) for the definition of $l_0$), and
inertial modes of $i_{l_0-|m|}$ have even (odd) parity when $l_0-|m|$ is an even (odd) integer.
We find that all the inertial modes in the table are stable.
The real part of $\bar\omega$ is insensitive to the Ekman number Ek but the imaginary part
depends on Ek and in general $\bar\omega_{\rm I}$ increases with increasing Ek.

\begin{figure}
\resizebox{0.31\columnwidth}{!}{
\includegraphics{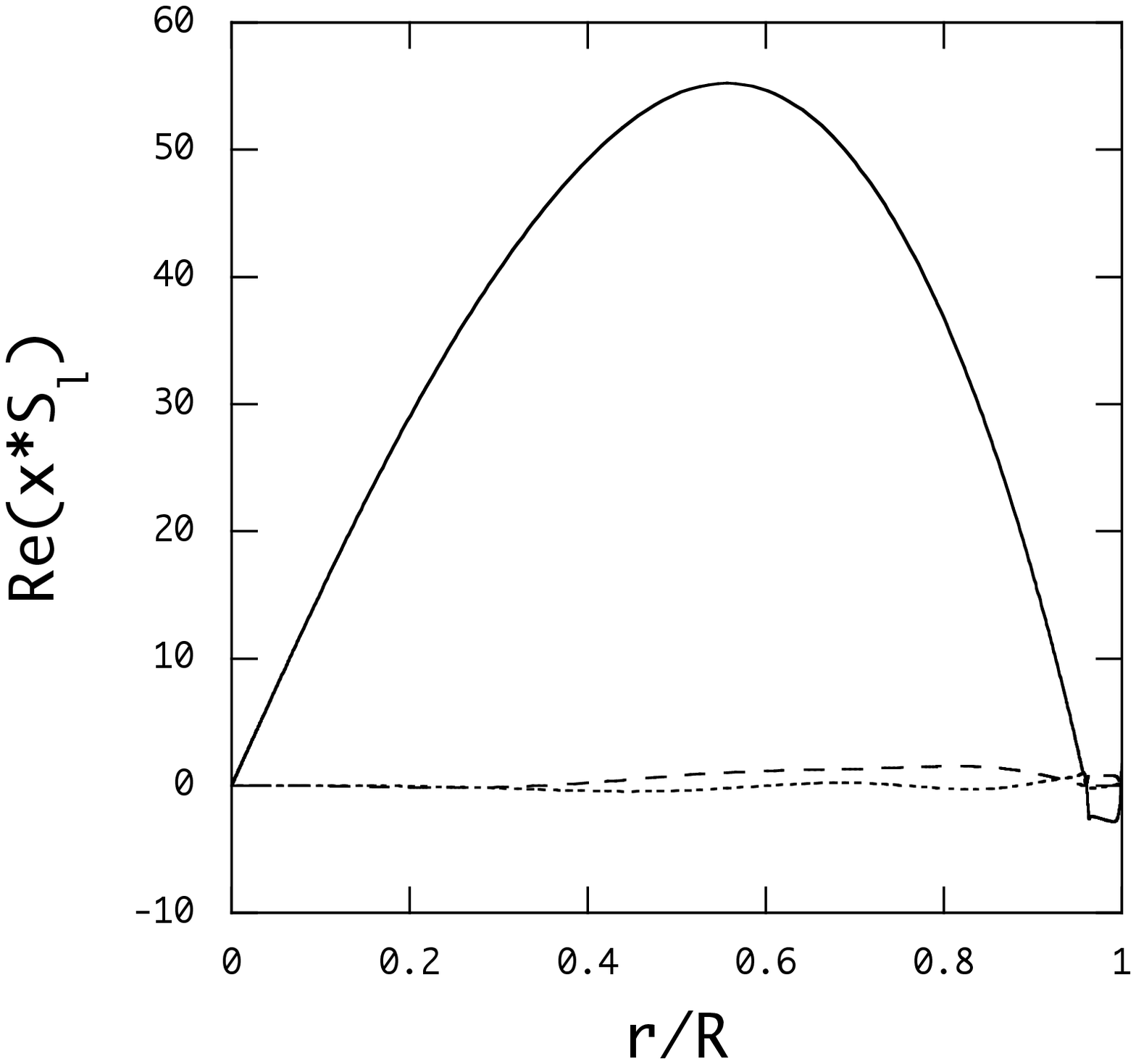}}
\hspace*{0.3cm}
\resizebox{0.31\columnwidth}{!}{
\includegraphics{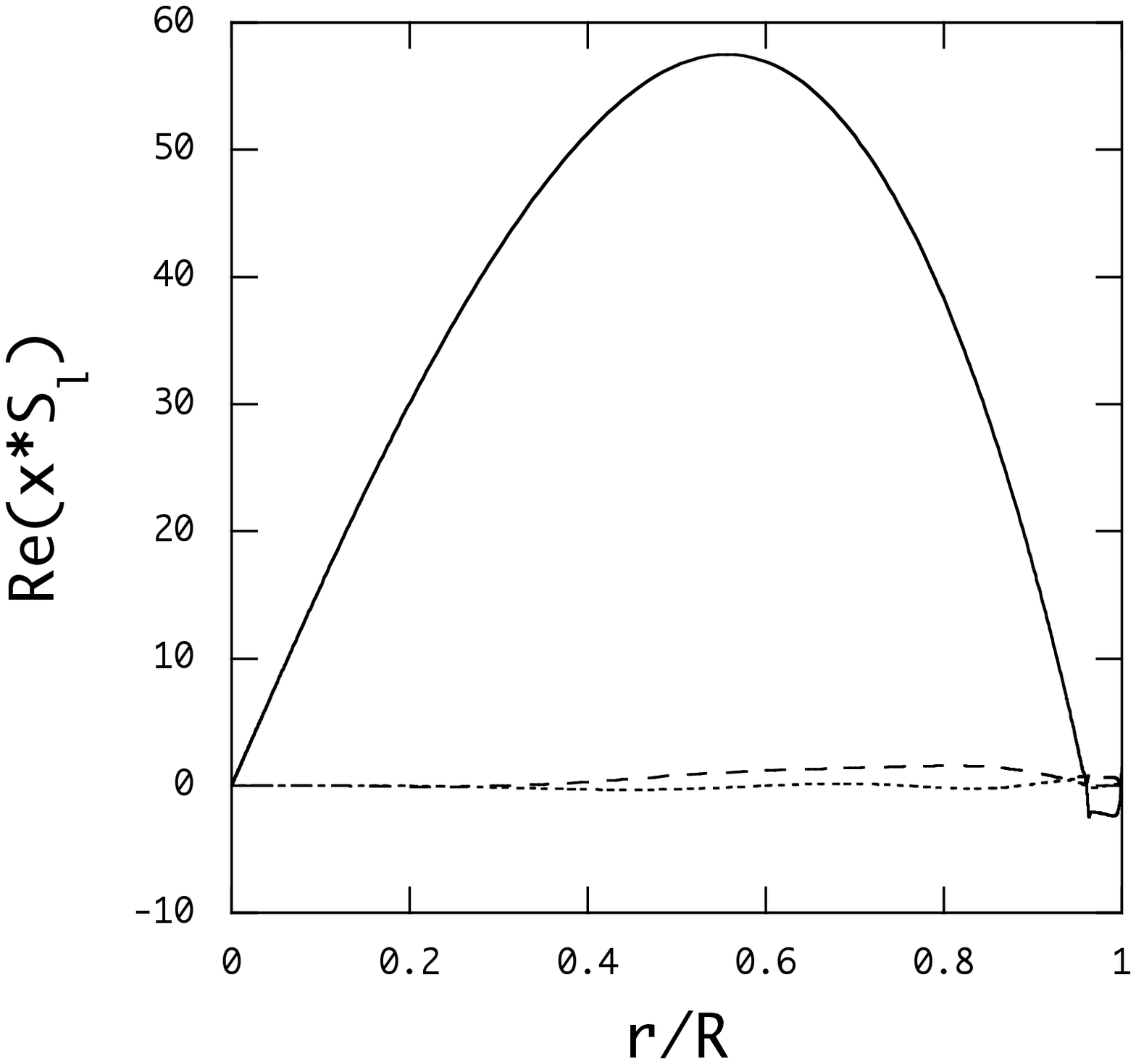}}
\hspace*{0.3cm}
\resizebox{0.31\columnwidth}{!}{
\includegraphics{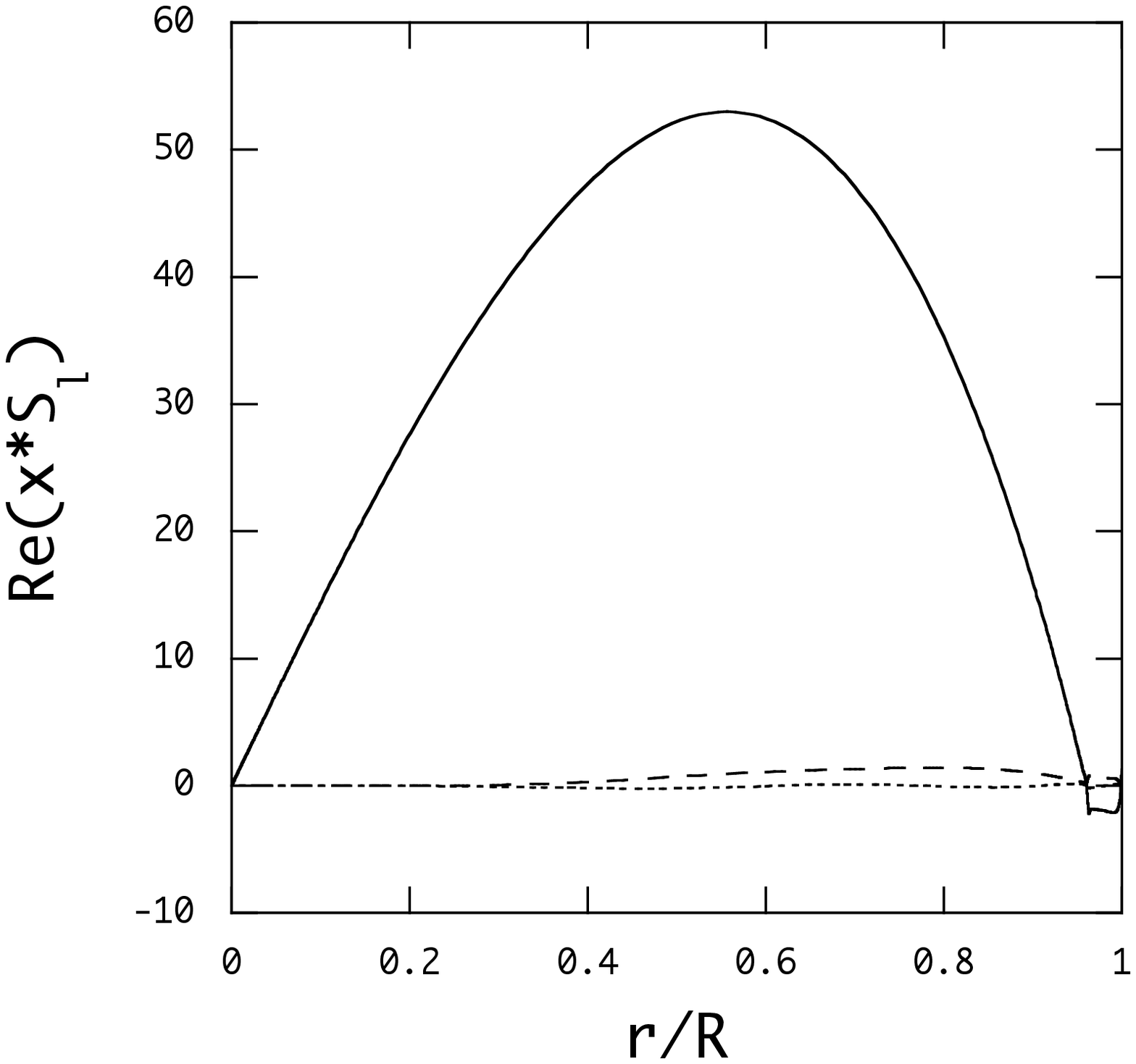}}
\caption{Expansion coefficients ${\rm Re}(xS_l)$ versus $x=r/R$ for
the $m=-2$ prograde inertial mode $i_2$ of $\bar\omega\approx0.055$ for
${\rm Ek}=10^{-9}$, $10^{-7}$ and $10^{-5}$, from left to right panels, where we use $\bar\Omega=0.1$ 
and $\tau_*=1$ day.
The solid, dashed, and dotted lines are for $l=2$, 4, and 6, respectively.
The normalization of the amplitudes is given by ${\rm Re}(S_{l_1})=1$ 
at the interface between the core and the envelope.
}
\label{fig:Sl_0055}
\end{figure}

\begin{figure}
\resizebox{0.31\columnwidth}{!}{
\includegraphics{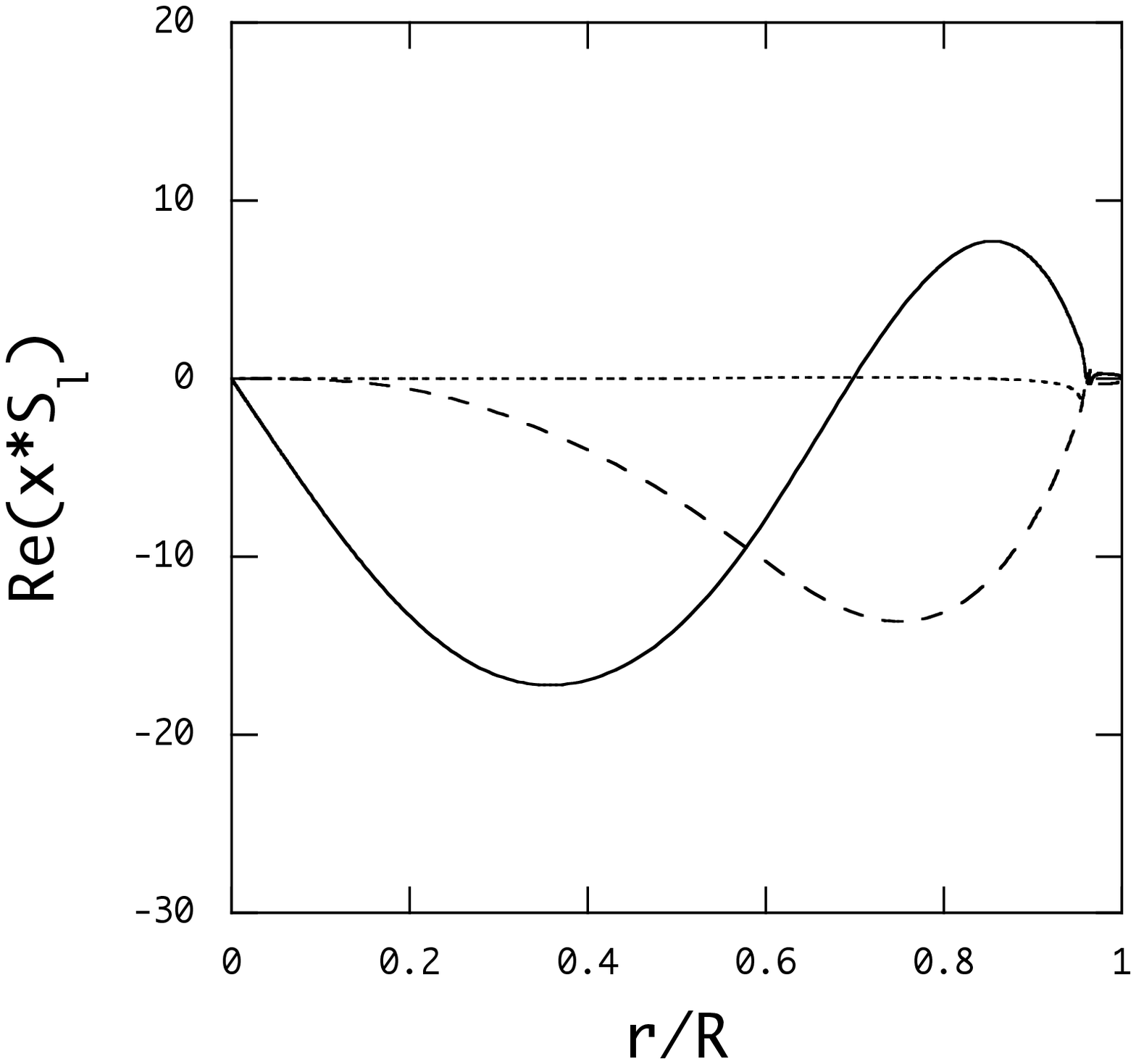}}
\hspace*{0.3cm}
\resizebox{0.31\columnwidth}{!}{
\includegraphics{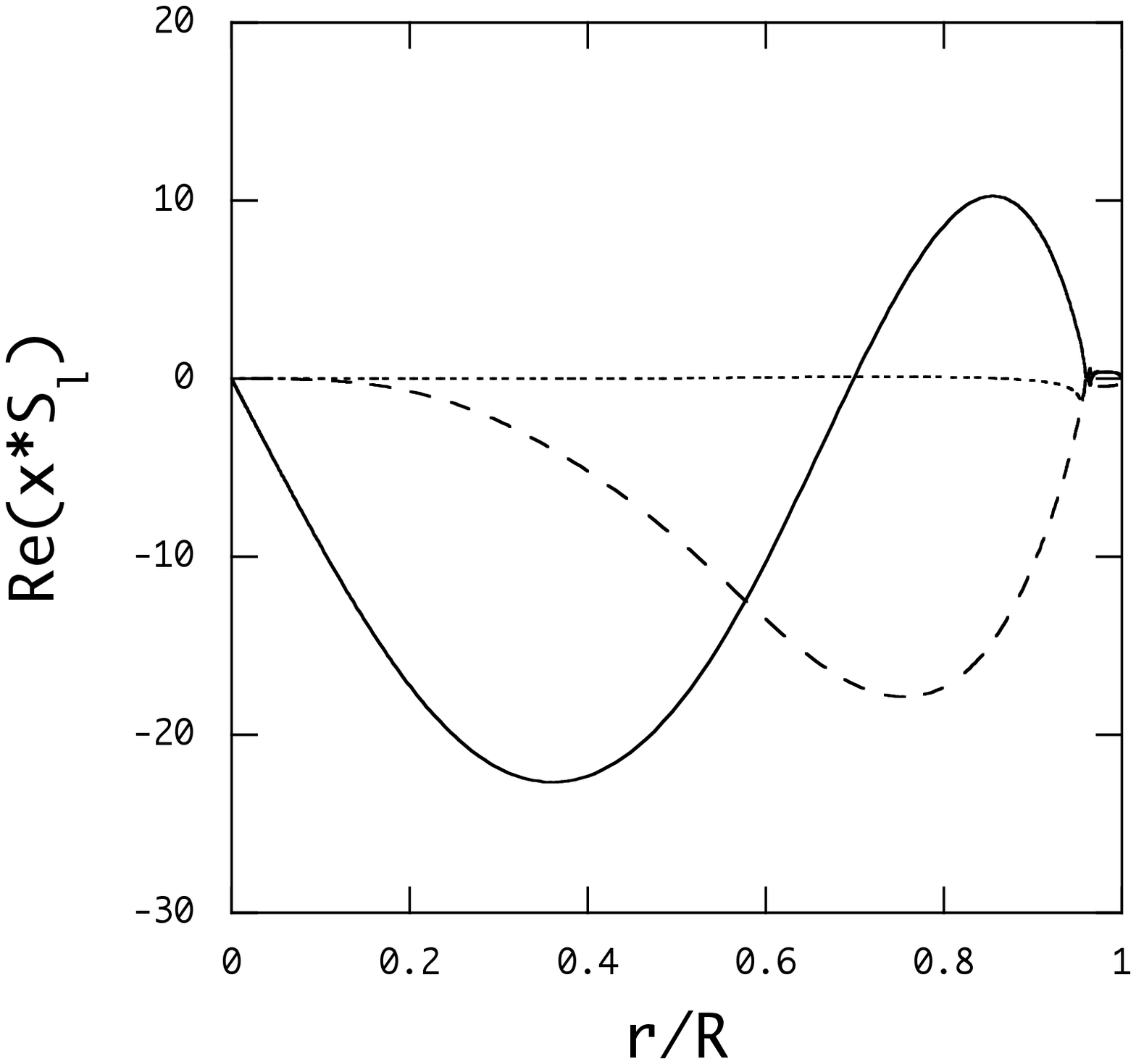}}
\hspace*{0.3cm}
\resizebox{0.31\columnwidth}{!}{
\includegraphics{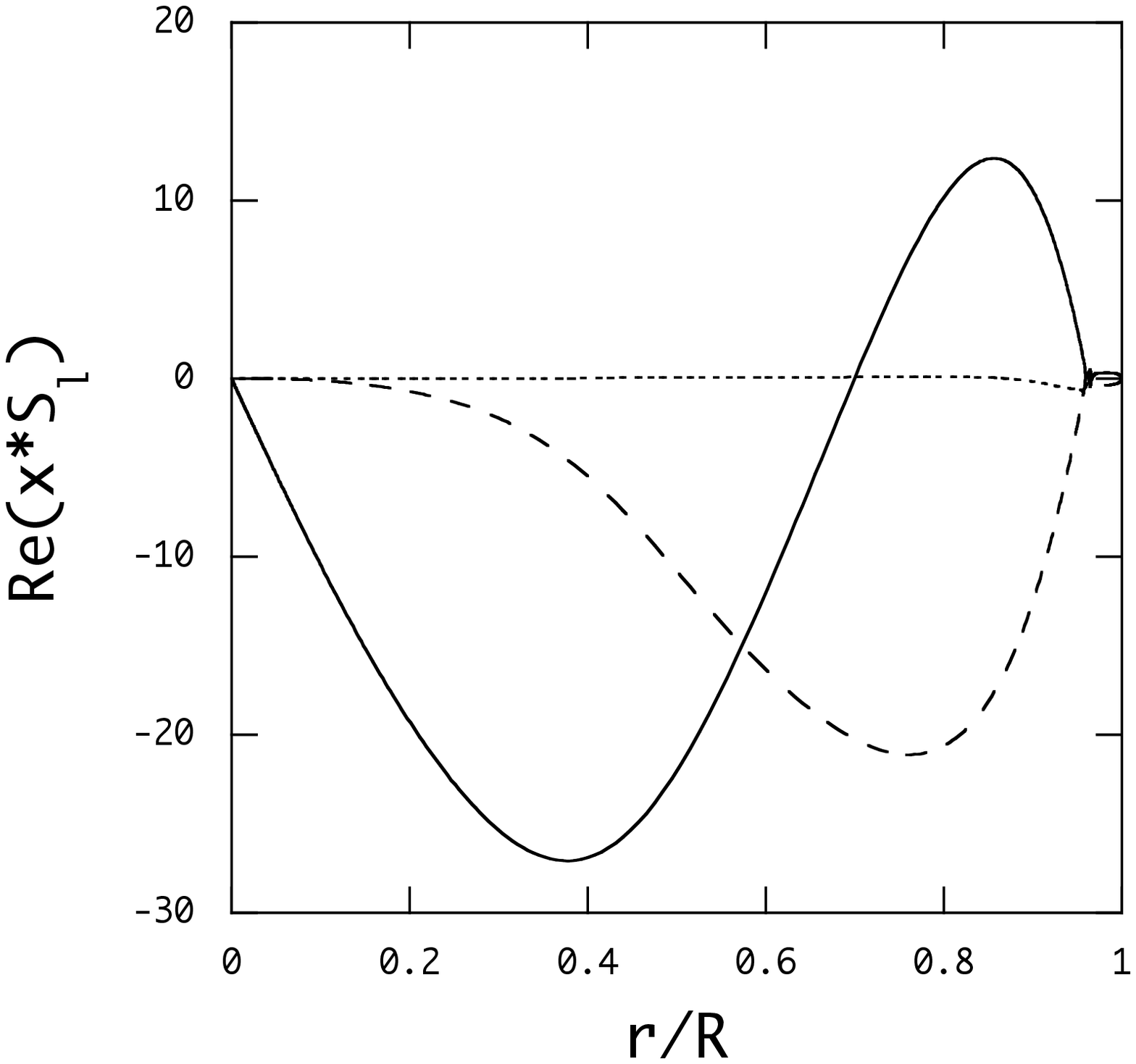}}
\caption{Same as Fig.\ref{fig:Sl_0055} but for a retrograde $i_4$ inertial mode of $\bar\omega\approx-0.086$.
}
\label{fig:Sl_m0085}
\end{figure}

In Fig. \ref{fig:Sl_0055} we plot the first few expansion coefficients $S_{l}$ 
of an $m=-2$ inertial mode of $\bar\omega\approx 0.055$, belonging to $i_2$ in Table 1, for 
${\rm Ek}=10^{-9}$, $10^{-7}$, and $10^{-5}$
where we use $\bar\Omega=0.1$ and $\tau_*=1$ day.
The coefficient $S_{l_1}$, which is well confined in the core, dominates other components and 
the eigenfunctions are insensitive to Ek.
Although we do not show any figures for the retrograde $i_2$ inertial mode of $\bar\omega\approx-0.11$,
we find quite similar properties of the expansion coefficients $S_l$ concerning their dependence on Ek.
Fig. \ref{fig:Sl_m0085} plot the first few expansion coefficients $S_l$ of an $i_4$ inertial mode of
$\bar\omega\approx -0.086$ for ${\rm Ek}=10^{-9}$, $10^{-7}$, and $10^{-5}$, where $\bar\Omega=0.1$ and $\tau_*=1$ day.
The first two expansion coefficients have dominating amplitudes, which are well confined in the core, and
the maximum amplitudes in the core increases as Ek increases.

\begin{table*}
\begin{center}
\caption{Complex eigenfrequency $\bar\omega=\bar\omega_{\rm R}+\rmi\bar\omega_{\rm I}$ of $m=-2$ inertial modes
of even parity in the viscous convective core for $\bar\Omega=0.1$ and $\tau_*=1$day. The figures $a(b)$ imply
$a\times10^b$. }
\begin{tabular}{@{}ccccccc}
\hline
    & \hspace*{2cm}${\rm Ek}=10^{-9}$ && \hspace*{2cm}${\rm Ek}=10^{-7}$ && \hspace*{2cm}${\rm Ek}=10^{-5}$  \\
\hline
 mode   & $\bar\omega_{\rm R}$ & $\bar\omega_{\rm I}$ & $\bar\omega_{\rm R}$ & $\bar\omega_{\rm I}$ & $\bar\omega_{\rm R}$ & $\bar\omega_{\rm I}$ \\
\hline
$i_2$  & $~~5.55(-2)$ & $~~1.11(-5)$ & $~~5.55(-2)$ & $~~1.33(-5)$ & $~~5.55(-2)$ & $~~6.40(-5)$  \\
$\cdots$  & $-1.10(-1)$ & $~~1.56(-6)$  & $-1.10(-1)$ & $~~2.33(-6)$  & $-1.10(-1)$ & $~~5.71(-5)$  \\
$i_4$  & $~~1.27(-1)$ & $~~1.00(-5)$ & $~~1.27(-1)$ & $~~2.28(-5)$  & $~~1.27(-1)$ & $~~1.84(-4)$  \\
$\cdots$  & $-1.52(-1)$ & $~~1.11(-5)$  & $-1.52(-1)$ & $~~2.58(-5)$  & $-1.52(-1)$ & $~~2.13(-4)$  \\
$\cdots$  & $~~2.74(-2)$ & $~~6.24(-6)$  & $~~2.74(-2)$ & $~~2.01(-5)$  & $~~2.74(-2)$ & $~~2.25(-4)$ \\
$\cdots$  & $-8.61(-2)$ & $~~3.29(-6)$  & $-8.61(-2)$ & $~~2.74(-5)$  & $-8.62(-2)$ & $~~2.20(-4)$  \\
\hline
\hline
\end{tabular}
\medskip
\end{center}
\end{table*}

\subsection{Tidal Torques}

The tidal responses are computed as forced oscillations by solving the set of inhomogeneous linear differential equations
with non-zero forcing terms $\pmb{\psi}$ and $\pmb{j}_*$ and the forcing frequency $\omega$ is treated as a real quantity.

\subsubsection{Calculating Tidal Torques}

The tidal torque ${\cal N}_0(m,\omega)$ on the planet for the azimuthal wavenumber $m$ and the forcing frequency $\omega$ 
may be given by
\be
{\cal N}_0(m,\omega)
=-{1\over 2}\int{\rm Re}\left({\partial\Phi_T\over\partial\phi}\rho'^*\right)dV
={m\over 2}\int{\rm Im}\left({\Phi_T}\rho'^*\right)dV,
\label{eq:tidaltorque00}
\ee
where  
the density perturbation $\rho'$ represents the tidal response to the perturbing forces $\Phi_T$ and $\epsilon'$ associated with $l=n=2$ and $m=-2$ or $m=+2$.

Since $\Phi_T$ given by equation (\ref{eq:phit0}) and $\epsilon'$ given by equations (\ref{eq:epsilonprime0}) 
and (\ref{eq:epsilonexpnasion}) are real quantities, in order to estimate the tidal torques driven by $\Phi_T$ and $\epsilon'$ we have to sum up the responses to
$\Phi_T^{(lm)}$ and its complex conjugate $(\Phi_T^{(lm)})^*$ for $\Phi_T$ and to
$J_*^{(nlm)}$ and its complex conjugate $(J_*^{(nlm)})^*$ for $\epsilon'$.
For given $m$ and $\omega$, if we let $\pmb{y}(m,\omega)\equiv(\pmb{y}_1,\pmb{Y}_2)$ denote the responses  
in the envelope to $\Phi_T^{(lm)}$ and $J_*^{(nlm)}$, 
the responses to the complex conjugates $(\Phi_T^{(lm)})^*$ and $(J_*^{(nlm)})^*$ may be given by
$\pmb{y}^*(m,\omega)$.
Using the relations given by $\beta_1(-\omega)=(\beta_1(\omega))^*$, $\beta_2(-\omega)=(\beta_2(\omega))^*$, and 
$\hat\Gamma_1(-\omega)=(\hat\Gamma_1(\omega))^*$, 
we find
\be
\pmb{y}^*(m,\omega)=\pmb{y}(-m,-\omega),
\ee
which leads to 
\be
{\cal N}\equiv{\cal N}_0(m,\omega) +{\cal N}_0(-m,-\omega)=2{\cal N}_0(m,\omega),
\ee
since $\rho'^*(-m,-\omega)=\rho'(m,\omega)$ and hence 
$-m{\rm Im}[\rho'^*(-m,-\omega)]=m{\rm Im}[\rho'^*(m,\omega)]$.
Similarly, 
for the responses $\pmb{z}_\alpha=(\pmb{z}_1,\pmb{Z}_2,\pmb{z}_3,\pmb{z}_6)$ and $\pmb{z}_\beta=(\pmb{z}_4,\pmb{z}_5)$ in the convective core,  
we find 
\be
\pmb{z}_\alpha^*(m,\omega)=\pmb{z}_\alpha(-m,-\omega), \quad 
\pmb{z}_\beta^*(m,\omega)=-\pmb{z}_\beta(-m,-\omega),
\ee
and 
\be
{\cal N}\equiv{\cal N}_0(m,\omega) +{\cal N}_0(-m,-\omega)=2{\cal N}_0(m,\omega).
\ee

Since $\tau_*$ represents the heat exchange timescale between the gas elements in the envelope,
radiative cooling becomes less efficient as $\tau_*$ increases and the responses in the envelope are
governed by
$\delta s\rightarrow  \epsilon'/(\rmi \omega T)$
as $\tau_*\rightarrow \infty$. 
In the core, $\epsilon '\approx 0$ and the responses tend to be adiabatic
as $\tau_*\rightarrow \infty$.
On the other hand, as $\tau_*$ decreases, radiative cooling becomes efficient in the envelope and the 
responses tend to be isothermal, that is, $T'\rightarrow 0$ as $\tau_*\rightarrow 0$.

\subsubsection{Tidal Torque versus the Forcing Frequency $\omega$ for constant $\Omega$}

\begin{figure}
\resizebox{0.33\columnwidth}{!}{
\includegraphics{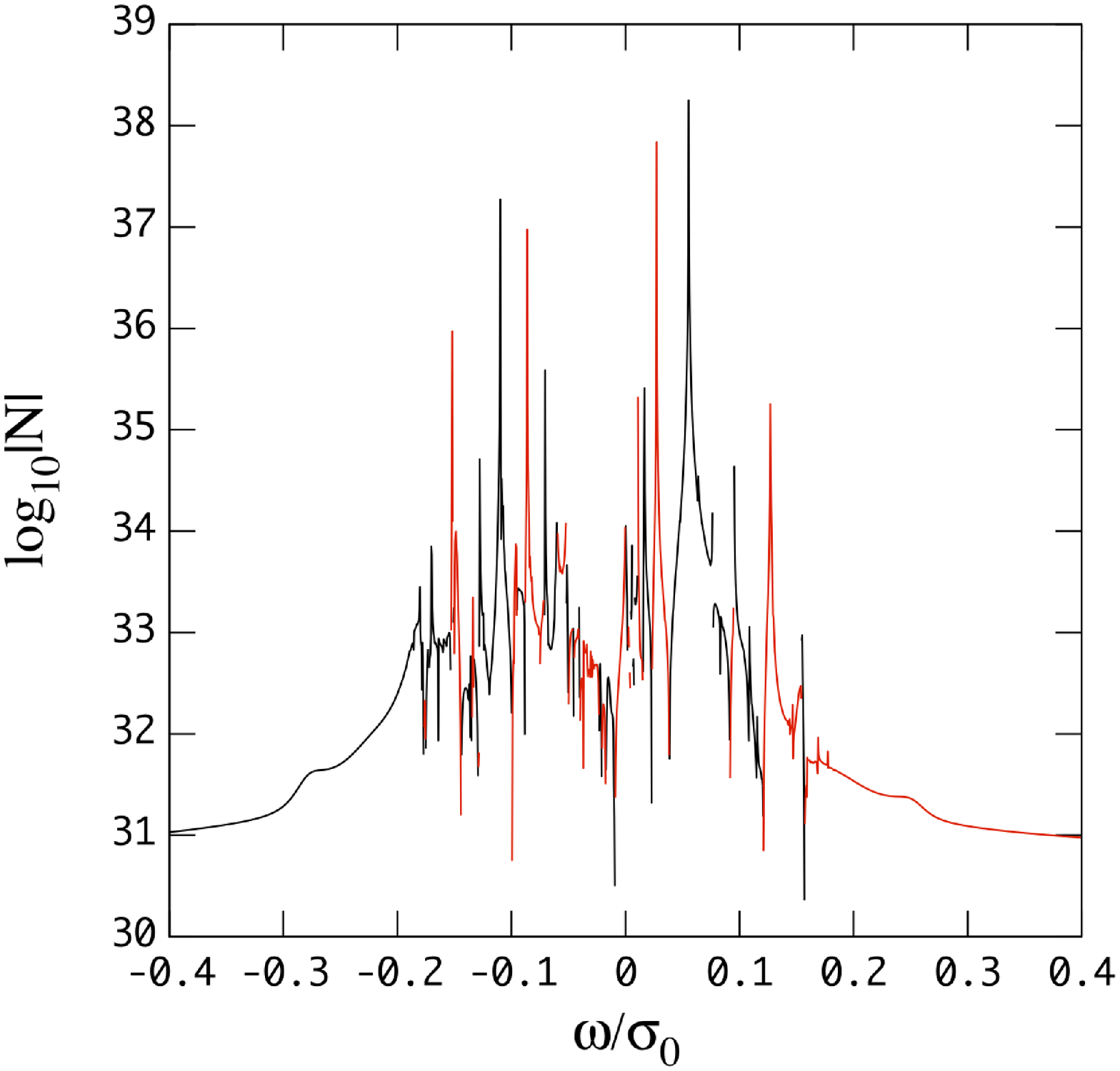}}
\resizebox{0.33\columnwidth}{!}{
\includegraphics{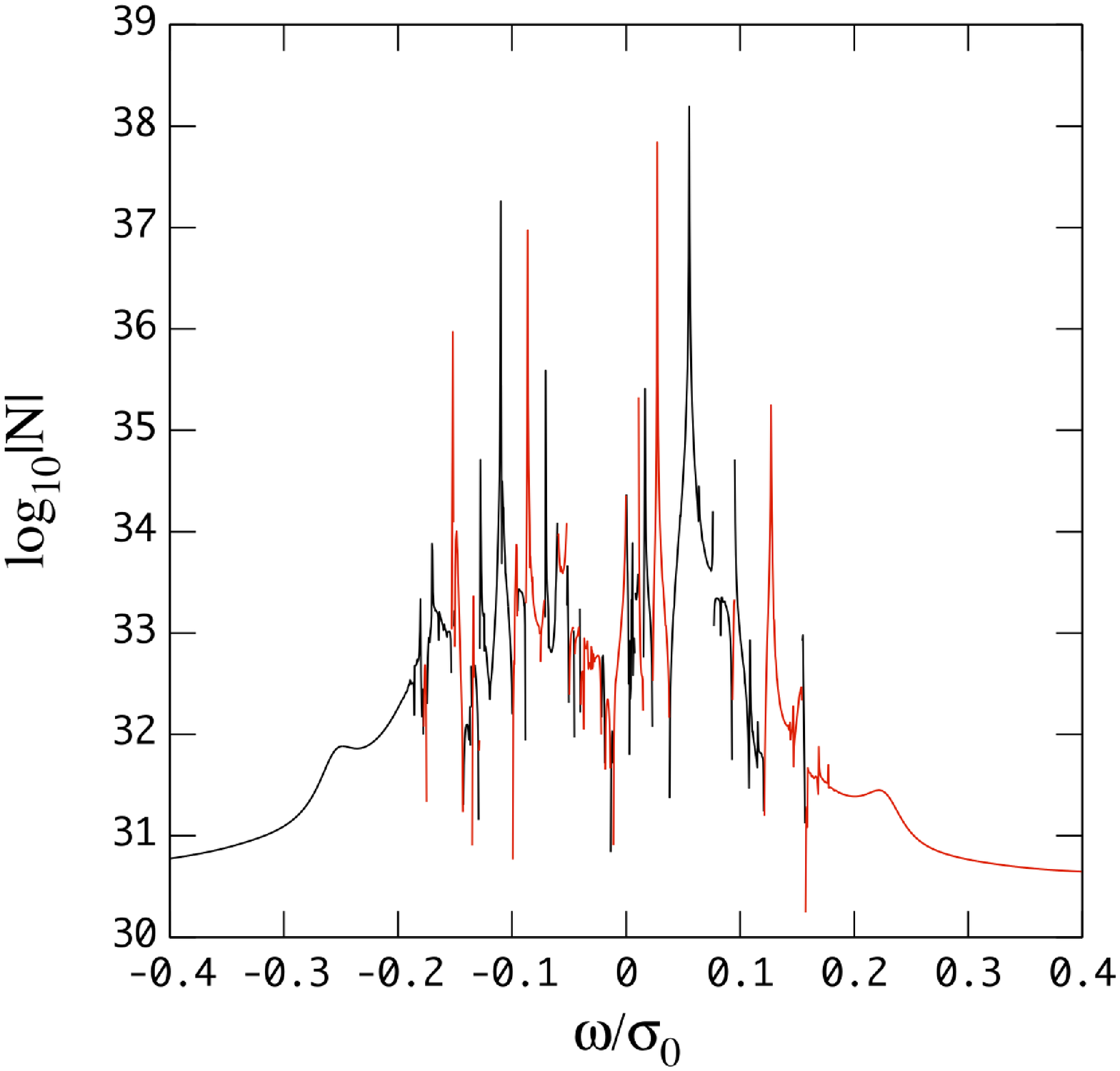}}
\resizebox{0.33\columnwidth}{!}{
\includegraphics{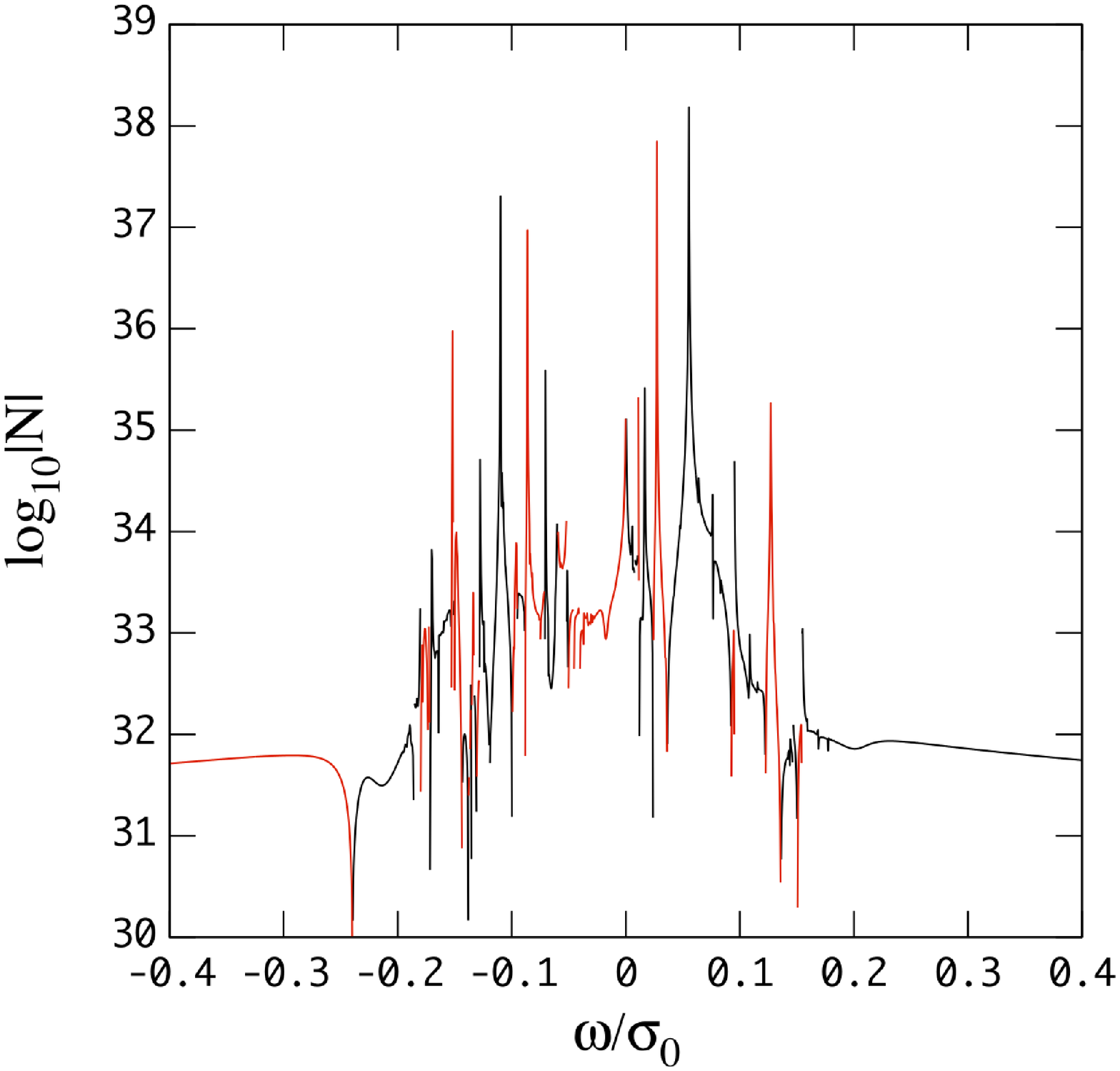}}
\caption{Tidal torques ${\cal N}$ in ergs
versus the forcing frequency $\omega/\sigma_0$ in the co-rotating frame
for $\tau_*=10$ day, 1 day, and 0.1 day, from left to right panels,
where ${\rm Ek}=10^{-7}$ and $\bar\Omega=0.1$ are assumed and the red (black) lines indicate positive
(negative) parts of the torque ${\cal N}$.
}
\label{fig:n_ekm7}
\end{figure}

\begin{figure}
\resizebox{0.33\columnwidth}{!}{
\includegraphics{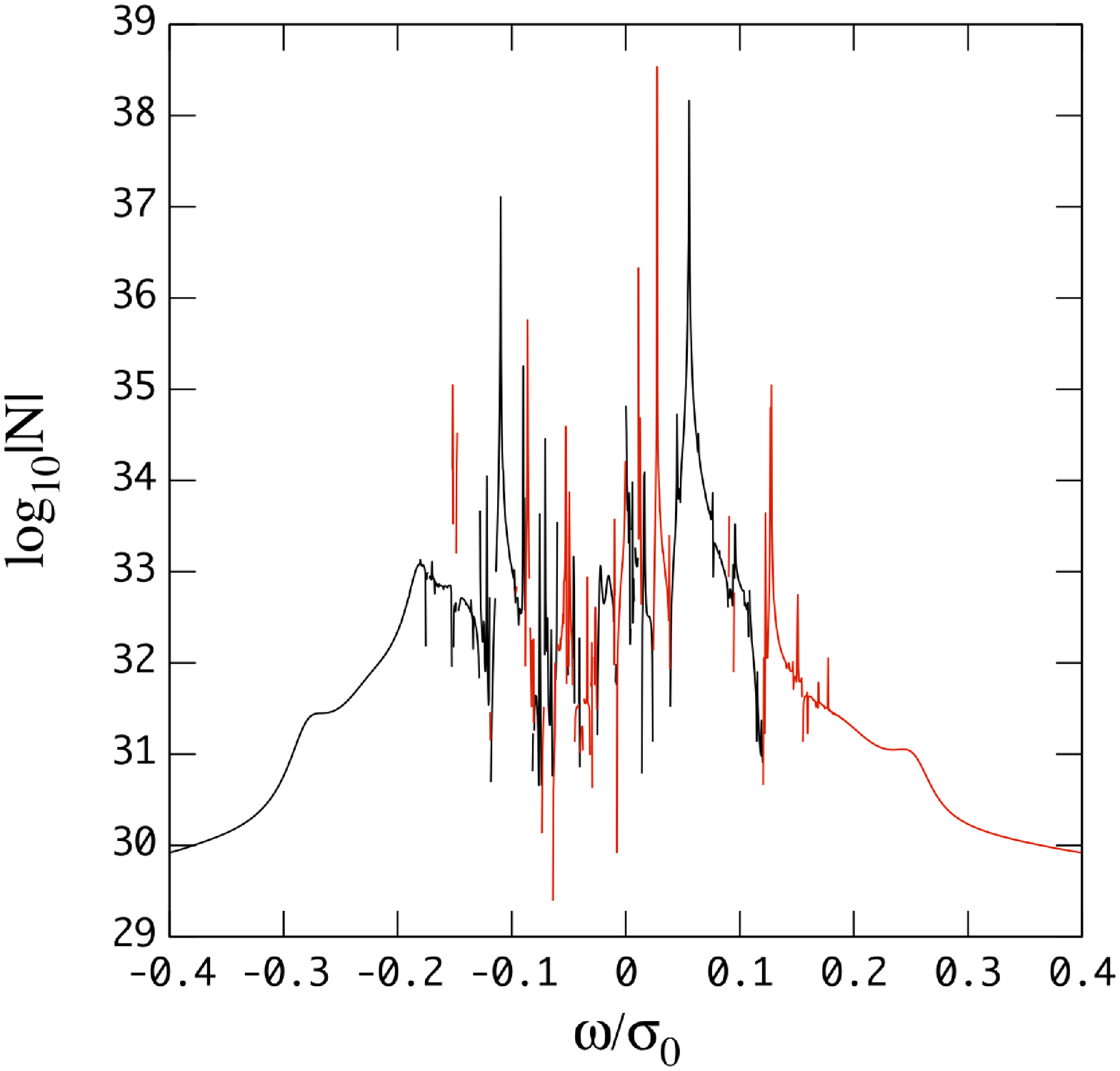}}
\resizebox{0.33\columnwidth}{!}{
\includegraphics{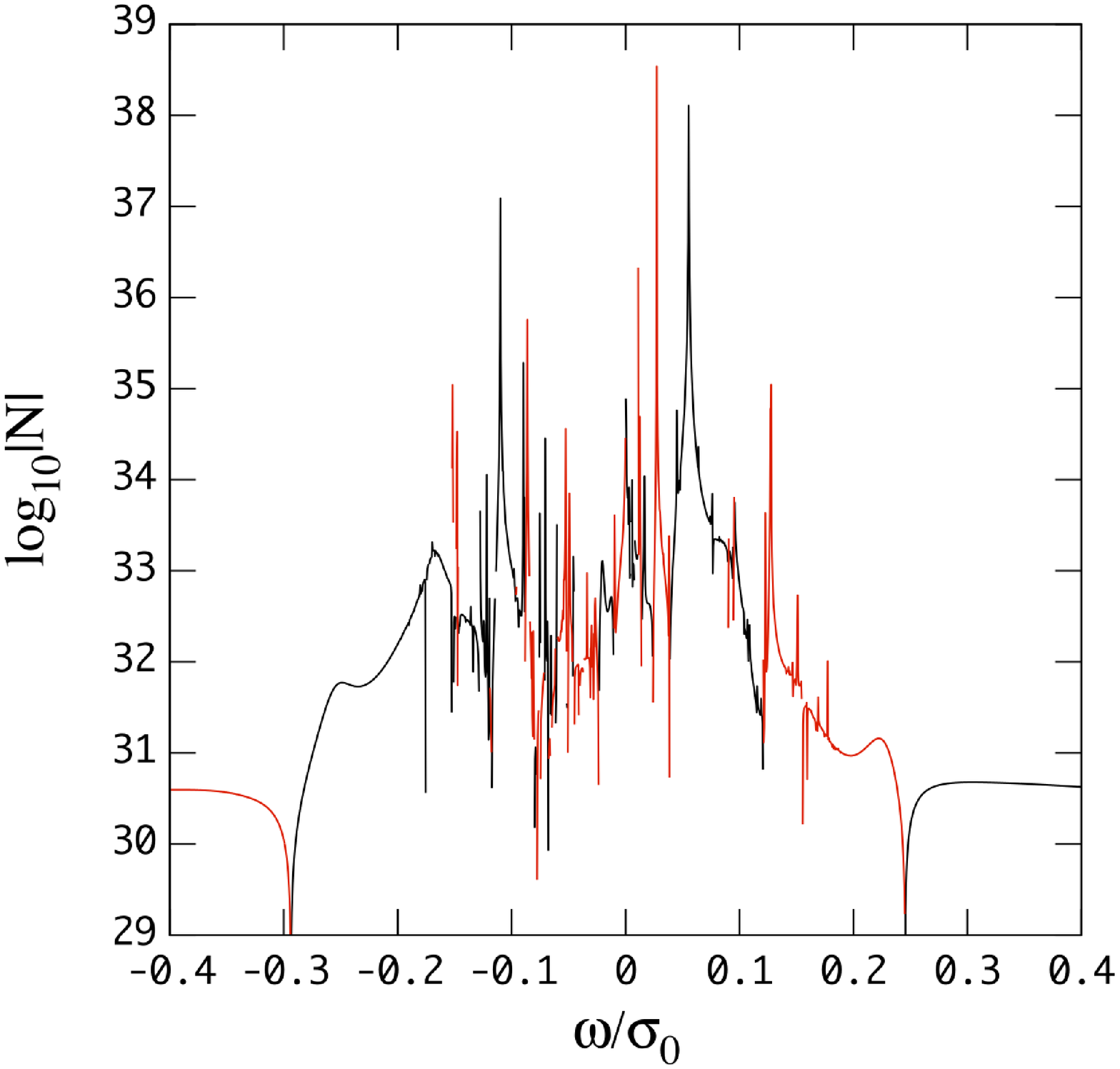}}
\resizebox{0.33\columnwidth}{!}{
\includegraphics{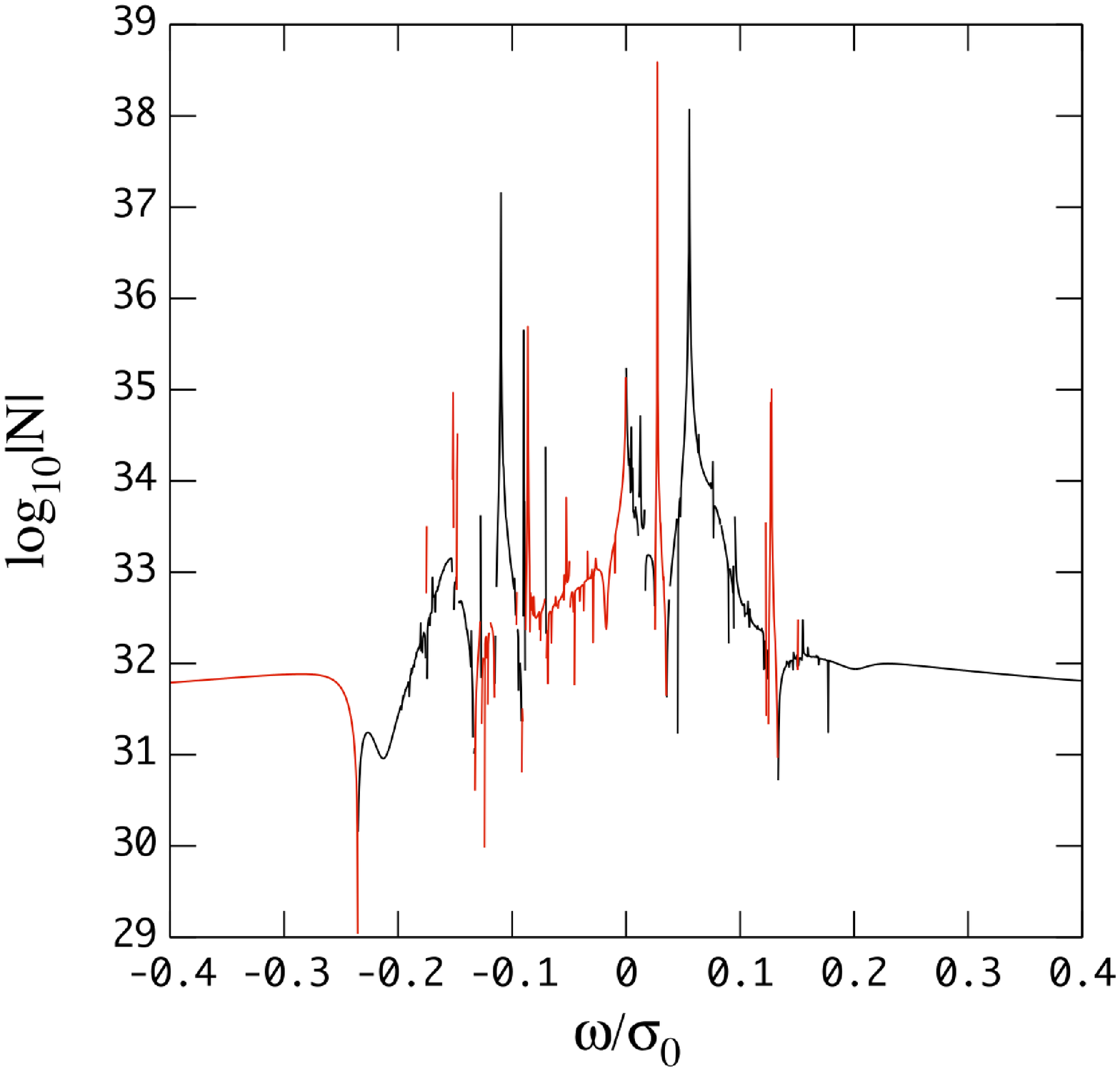}}
\caption{Same as Fig.\ref{fig:n_ekm7} but for Ek$=10^{-9}$.
}
\label{fig:n_ekm9}
\end{figure}

\begin{figure}
\resizebox{0.33\columnwidth}{!}{
\includegraphics{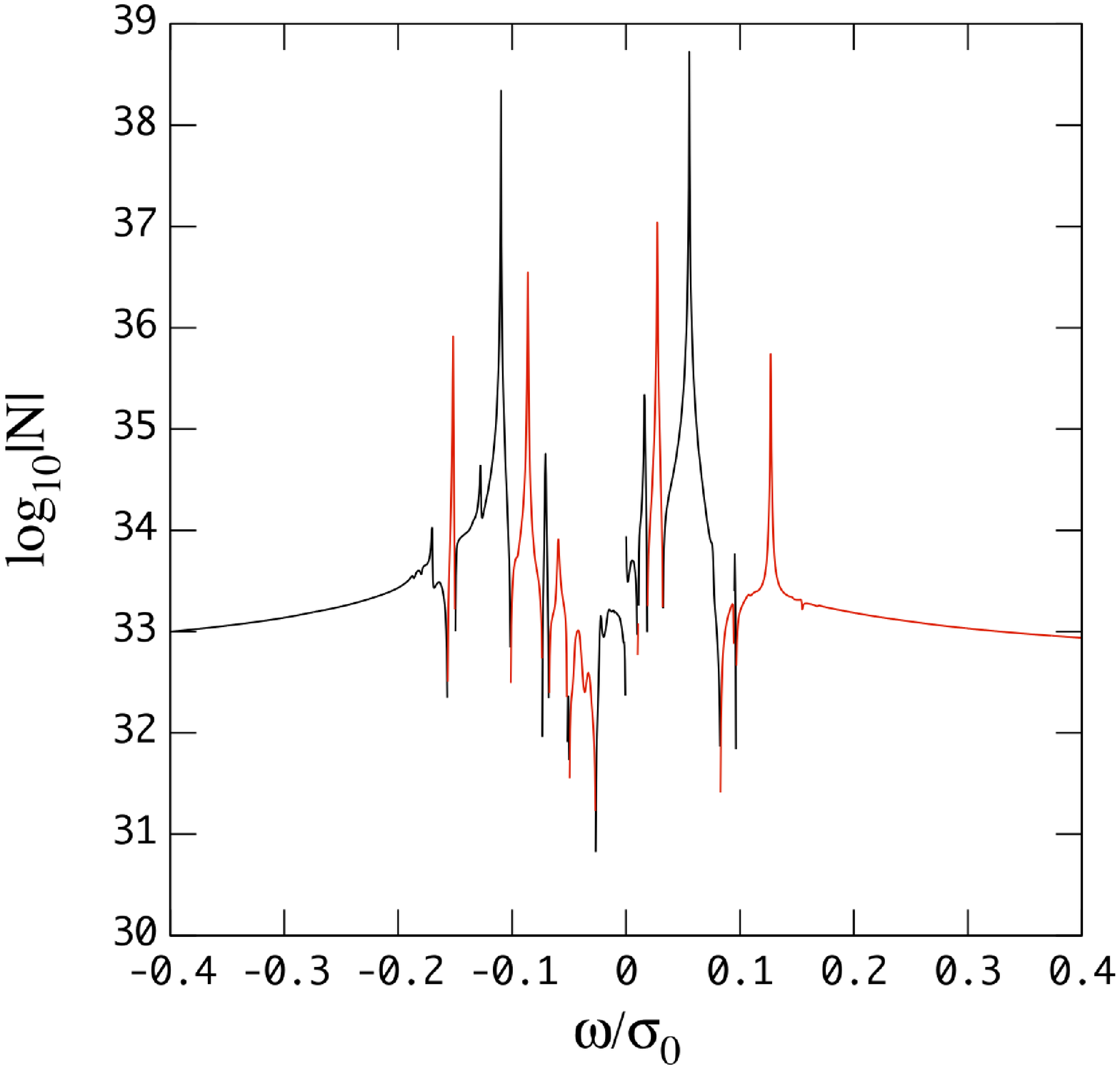}}
\resizebox{0.33\columnwidth}{!}{
\includegraphics{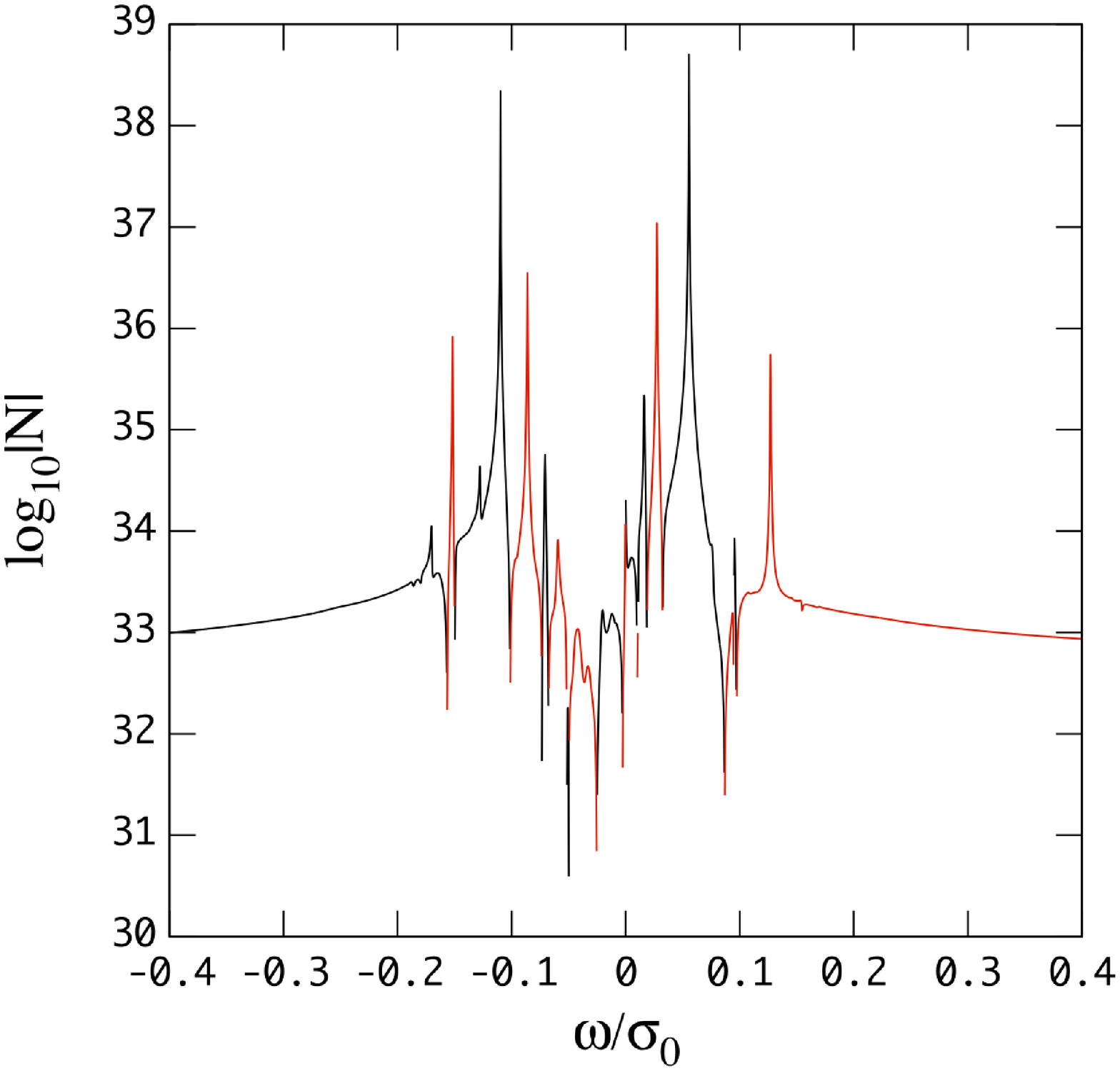}}
\resizebox{0.33\columnwidth}{!}{
\includegraphics{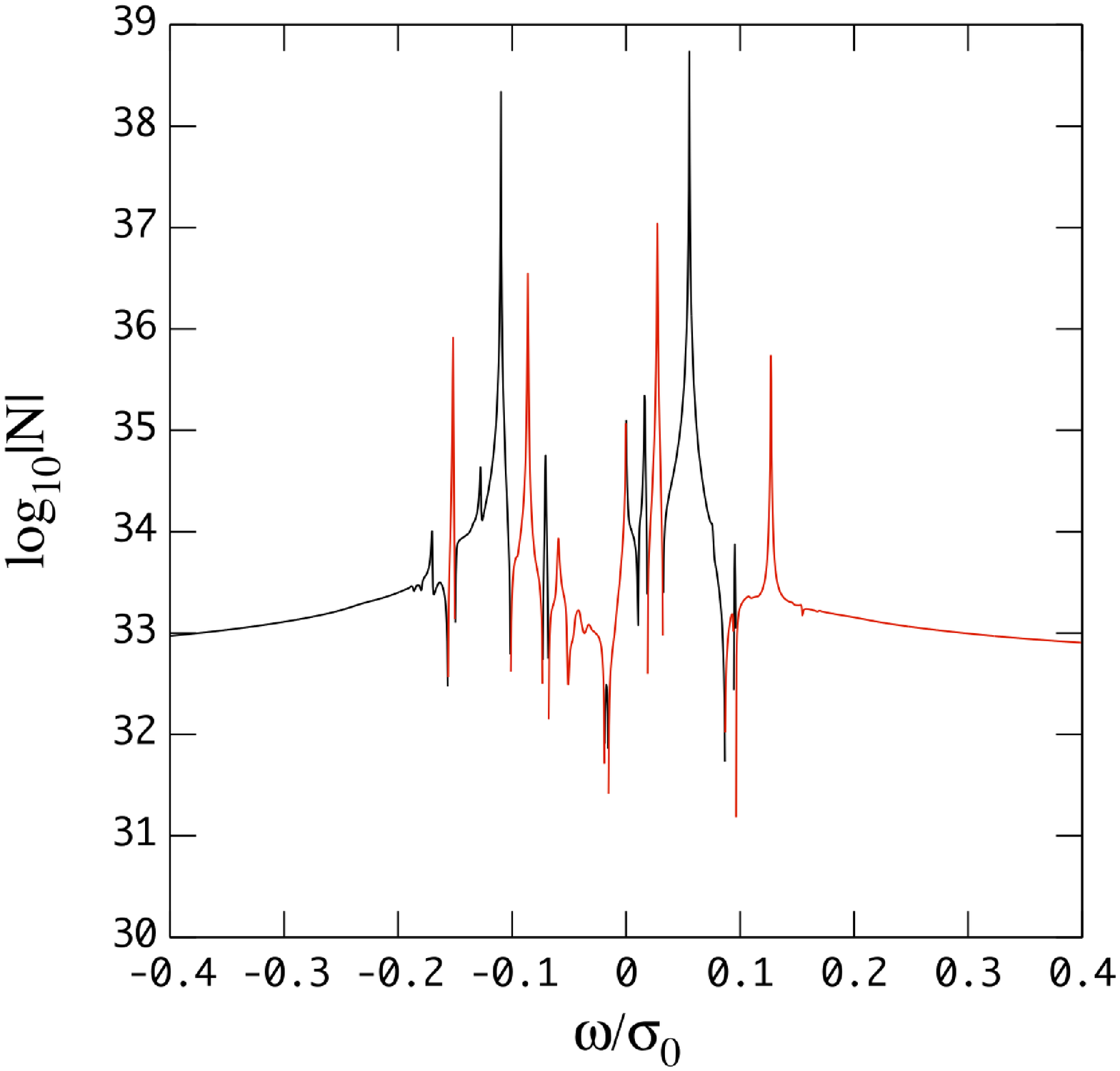}}
\caption{Same as Fig.\ref{fig:n_ekm7} but for Ek$=10^{-5}$.
}
\label{fig:n_ekm5}
\end{figure}

Fig. \ref{fig:n_ekm7} plots the tidal torque ${\cal N}$ as a function of 
the forcing frequency $\bar\omega$ for $\tau_*=10$ day, 1 day, and 0.1 day
where we use ${\rm Ek}=10^{-7}$ and $\bar\Omega=0.1$ and
the red (black) lines indicate positive (negative) parts of the torque ${\cal N}$.
In our convection, positive (negative) $\bar\omega$ corresponds to the prograde (retrograde) forcing in the co-rotating frame of the planet for $m<0$ and $\bar\Omega>0$.
To attain synchronization between the orbital motion and spin of the planets, the tidal torques
${\cal N}$ need to be positive (negative) for prograde (retrograde) forcing.
As computed by Lee \& Murakami (2019) for the same hot Jupiter models, 
$g$-modes in the envelope have frequencies $|\bar\omega|\lesssim 0.1$ for $\bar\Omega=0.1$ and $\tau_*=1$ day.
Frequency resonances of the tidal forcing with the envelope $g$-modes produce broad resonance peaks of ${\cal N}$,
which are disturbed by many sharp peaks produced by resonance with core inertial modes.
For $\bar\Omega=0.1$, core inertial modes appear in the inertial range $|\bar\omega|\le 2\bar\Omega=0.2$.
As a result of resonance between the forcing $\bar\omega$ and inertial modes in the core,
there appear prominent peaks of ${\cal N}$ at the forcing frequencies corresponding to 
the $i_2$ and $i_4$ inertial modes tabulated in Table 1.
For ${\rm Ek}=10^{-7}$, the most prominent peak of $\cal N$ is due to the resonance with the prograde $i_2$ inertial mode of $\bar\omega\approx 0.055$ (see Table 1), and
the resonances with the retrograde $i_2$ inertial mode of $\bar\omega\approx-0.11$ and with
the $i_4$ inertial modes also produce very high peaks of ${\cal N}$.
In addition to these prominent resonance peaks, there appear numerous minor peaks produced by resonance with inertial modes $i_n$ with $n\ge 6$.
Inertial modes $i_n$ with large $n$ in general have very short wave lengths in the core.
It is interesting to note that the sign of the peaks of ${\cal N}$ produced by the $i_2$ inertial modes is negative
and that by the $i_4$ modes is positive, irrespective of $\tau_*$.
Since the resonance with the prograde $i_2$ mode produces a negative peak of ${\cal N}$,
we may suggest a possibility that 
if the tidal forcing gets into resonance with the prograde $i_2$ inertial mode,
the negative ${\cal N}$ in the resonance will hamper the process of 
synchronization between the orbital motion and spin of the planets.

In the low frequency region $|\bar\omega|\gtrsim 0.2$ for $\bar\Omega=0.1$, 
no resonance features associated with inertial modes appear as shown by 
Fig. \ref{fig:n_ekm7}.
This figure also shows that the sign of ${\cal N}$ outside this inertial range depends on $\tau_*$.
For $\tau_*\gtrsim 1$ day, ${\cal N}$ is negative on the retrograde side and it is positive on the prograde side,
suggesting that the tidal forcing contributes to synchronization between the orbital motion and spin of the planets.
For $\tau_*\lesssim 0.1$ day, however, ${\cal N}$ is positive on the retrograde side and negative on the prograde side.
This dependence of ${\cal N}$ on $\tau_*$ may suggest that dissipative contributions from the viscous core to ${\cal N}$ are comparable to or smaller than thermal contributions from the radiative envelope for ${\rm Ek}\sim 10^{-7}$.

Similar dependence of ${\cal N}$ outside the inertial range on $\tau_*$ is found for ${\rm Ek}=10^{-9}$ as shown by Fig. \ref{fig:n_ekm9}.
We may also find that gross properties of ${\cal N}$ in the inertial range for ${\rm Ek}=10^{-9}$
are quite similar to those for ${\rm Ek}=10^{-7}$, suggesting that thermal contributions from the radiative
envelope are dominating to determine $\cal N$ if Ek is small enough.

If Ek is large enough, on the other hand, ${\cal N}$ outside the inertial range is always
negative on the retrograde side and positive on the prograde side as shown by Fig. \ref{fig:n_ekm5}
for Ek$=10^{-5}$ and its magnitude is proportional to Ek as suggested by
comparing ${\cal N}$'s at $\bar\omega=0.4$ in Figs. \ref{fig:n_ekm7} and \ref{fig:n_ekm5} for $\tau_*\gtrsim 1$ day.
We also note that as Ek increases, the number of high resonance peaks that appear in the inertial range 
decrease.
This is because $\bar\omega_{\rm I}$ of free inertial modes in the core increases with increasing Ek
and the widths of the resonance peaks due to the inertial modes will widen.
If $\bar\omega_{\rm I}$ becomes comparable to $|\bar\omega_{\rm R}|$, the peaks will be smoothed out.

\begin{figure}
\resizebox{0.32\columnwidth}{!}{
\includegraphics{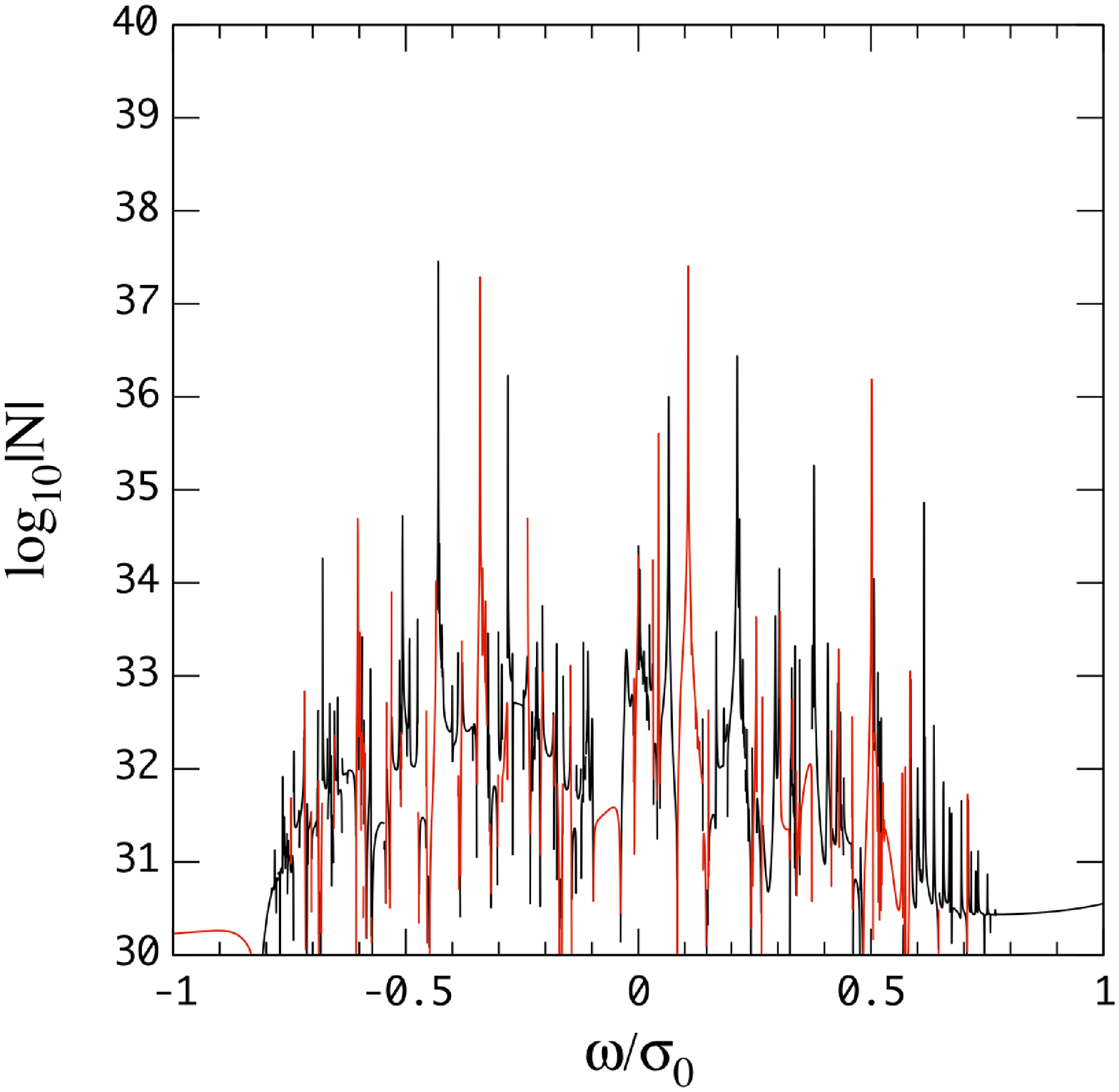}}
\hspace*{0.3cm}
\resizebox{0.32\columnwidth}{!}{
\includegraphics{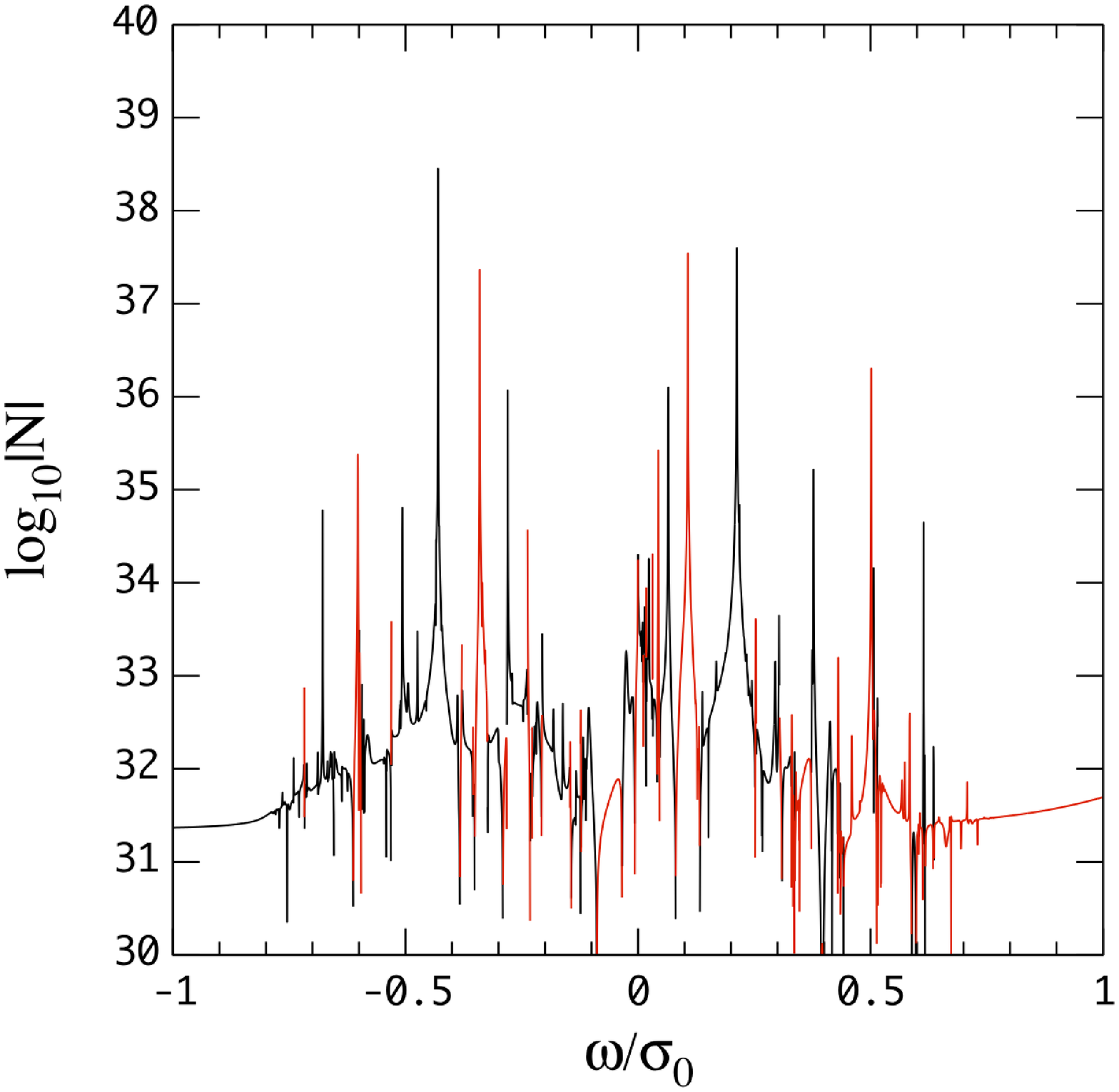}}
\hspace*{0.3cm}
\resizebox{0.32\columnwidth}{!}{
\includegraphics{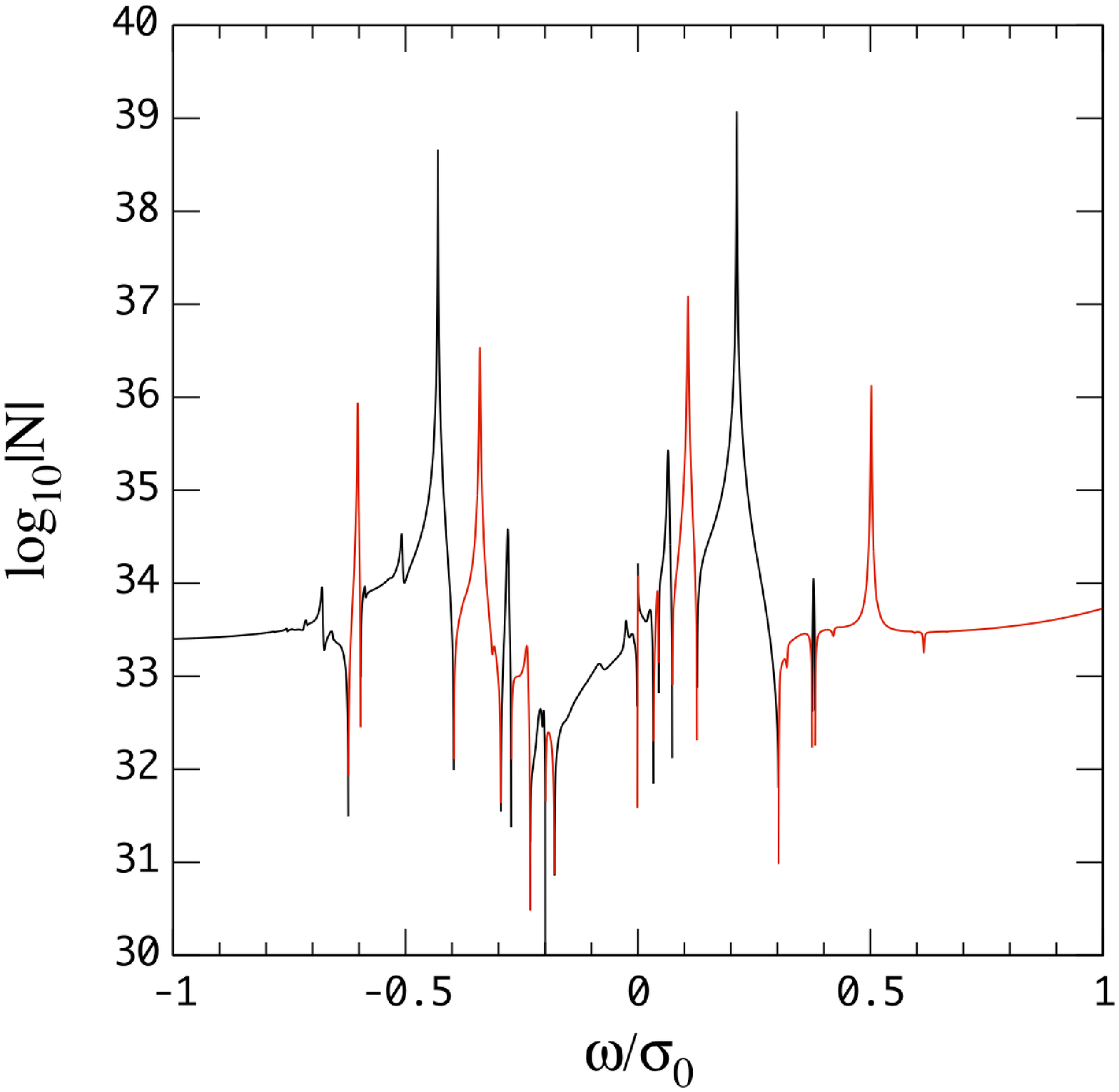}}
\caption{Tidal torque ${\cal N}$ versus the forcing frequency $\bar\omega$
for EK$=10^{-9}$, $10^{-7}$, and $10^{-5}$, from left to right panels, where we assume $\bar\Omega=0.4$ and 
$\tau_*=1$ day.
}
\label{fig:tau1d10omega04}
\end{figure}

For hot Jupiters rapidly rotating at $\bar\Omega=0.4$, for example,
the inertial range of the forcing frequency extends to $|\bar\omega|\le 0.8$.
In Fig. \ref{fig:tau1d10omega04},
we plot the tidal torque ${\cal N}$ for $\bar\Omega=0.4$ as a function of $\bar\omega$ 
for ${\rm Ek}=10^{-9}$, $10^{-7}$, and $10^{-5}$ where we use $\tau_*=1$ day.
We find numerous resonance peaks of ${\cal N}$ due to the core inertial modes, particularly for Ek$=10^{-9}$.
The high peaks are 
produced by resonance with the $i_2$ and $i_4$ inertial modes and
the signs of the $i_2$ peaks are negative and those of the $i_4$ peaks are positive, which 
is the same as the result obtained for $\bar\Omega=0.1$.

\subsubsection{Tidal Torque versus the Forcing Period $\tau_{\rm tide}$ for constant $\Omega_{\rm orb}$}

\begin{figure}
\resizebox{0.45\columnwidth}{!}{
\includegraphics{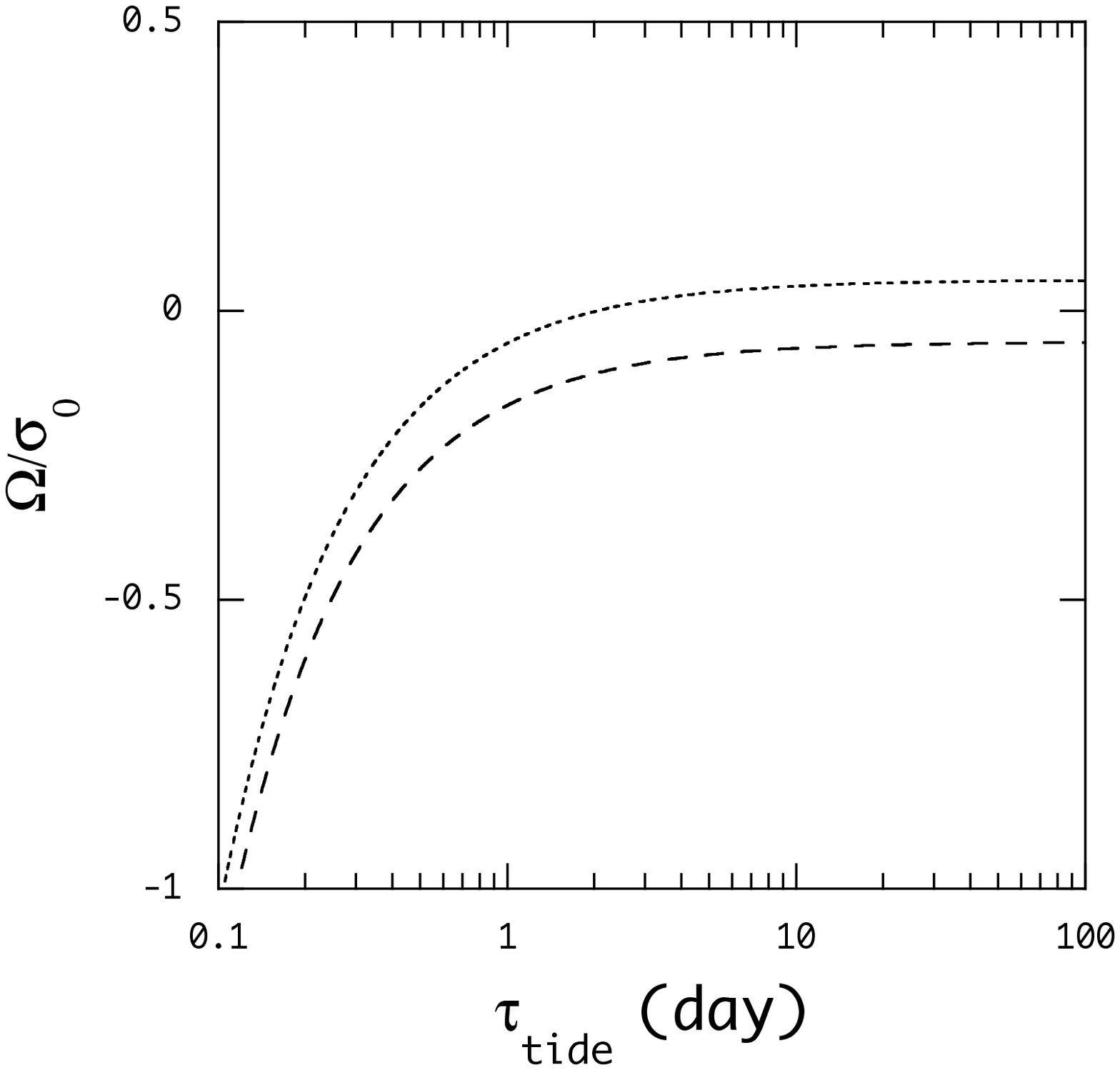}}
\hspace*{0.5cm}
\resizebox{0.45\columnwidth}{!}{
\includegraphics{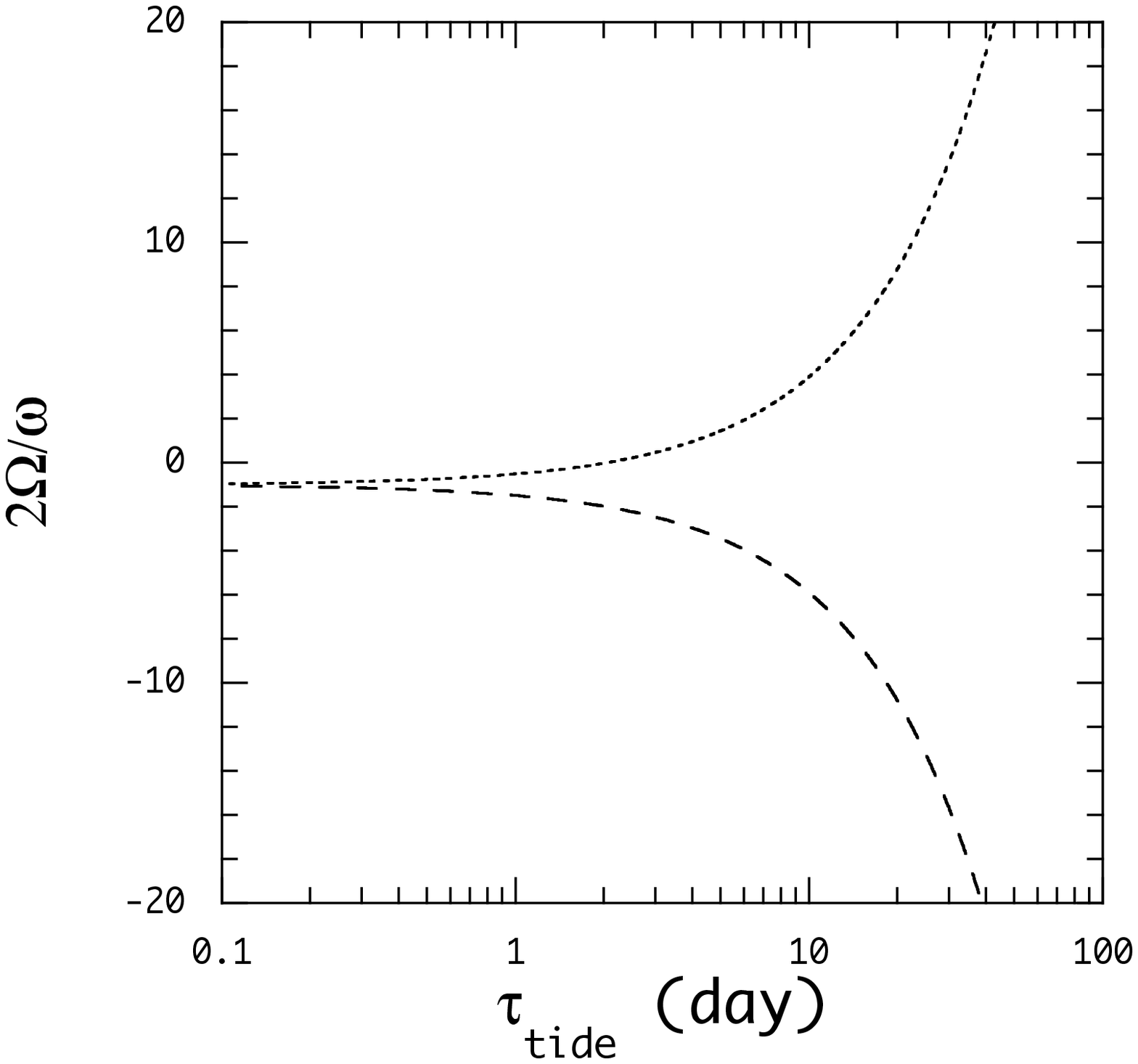}}
\caption{Angular rotation speed $\bar\Omega$ (left panel) and the spin parameter $2\Omega/\omega$
(right panel) as a function of the forcing period $\tau_{\rm tide}$ in day, where the dotted lines and
dashed lines are for $\bar\Omega_{\rm orb}=0.0537$ and $\bar\Omega_{\rm orb}=-0.0537$, respectively.
}
\label{fig:f26}
\end{figure}

Assuming that the spin frequency $\Omega$ of the planet changes with the forcing period 
$\tau_{\rm tide}\equiv 2\pi/\omega$ for given $\Omega_{\rm orb}$, that is,
\be
\Omega=\Omega_{\rm orb}-\pi/\tau_{\rm tide},
\ee
we calculate the tidal torques as a function of $\tau_{\rm tide}$ for $\bar\Omega_{\rm orb}=\pm0.0537$ 
for the forcing period ranging from 0.1 day to 100 day.
Fig. \ref{fig:f26} plots
the rotation frequency $\bar\Omega$ (left panel) and the spin parameter $2\Omega/\omega$
(right panel) versus $\tau_{\rm tide}$ where the dotted and the dashed lines in the panels are for $\bar\Omega_{\rm orb}=0.0537$ and $\bar\Omega_{\rm orb}=-0.0537$, respectively.
At the forcing period $\tau_{\rm tide}\sim0.1$ day, the forcing frequency is $\bar\omega\sim 2$, which
is close to the frequencies of low radial order $p$-modes of the planets, and
the rotation frequency is $\bar\Omega\sim1$, that is, almost the break-up velocity of the planets.
For $\bar\Omega_{\rm orb}=\pm0.0537$, $\bar\Omega$ is negative at $\tau_{\rm tide}\sim0.1$ day 
and monotonously increases with increasing $\tau_{\rm tide}$.
As $\tau_{\rm tide}$ increases, $\bar\Omega$ stays negative for $\bar\Omega_{\rm orb}=-0.0537$, but  
for $\bar\Omega_{\rm orb}=0.0537$, $\bar\Omega$ vanishes at $\tau_{\rm tide}\approx 2$ day and changes its sign from being
negative to positive.
The forcing period $\tau_{\rm tide}$ is in the inertial range 
when $\tau_{\rm tide}\gtrsim4$ day for $\bar\Omega_{\rm orb}=0.0537$ and $\tau_{\rm tide}\gtrsim 0.1$ day 
for $\bar\Omega_{\rm orb}=-0.0537$.

\begin{figure}
\resizebox{0.45\columnwidth}{!}{
\includegraphics{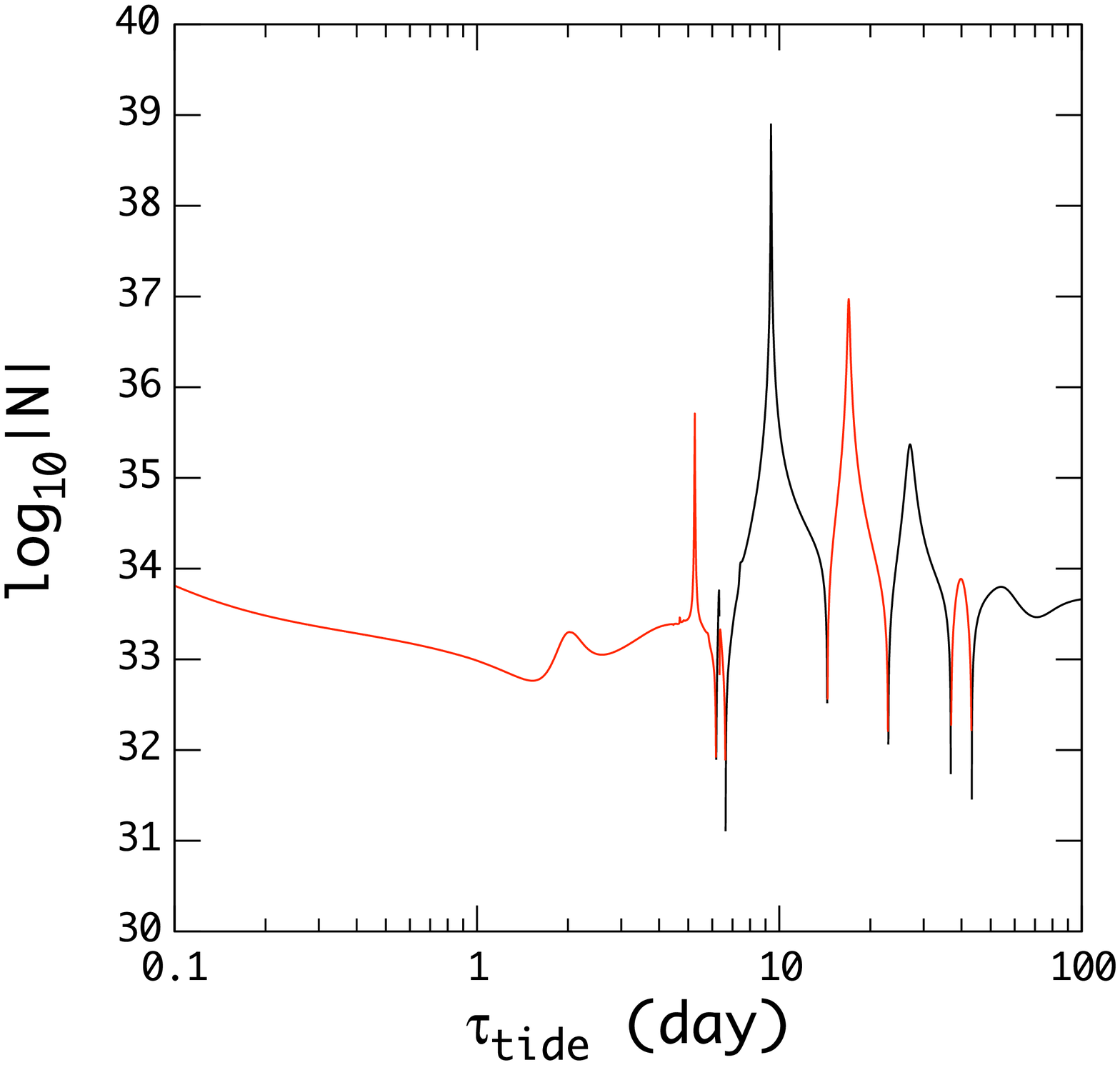}}
\hspace*{1cm}
\resizebox{0.45\columnwidth}{!}{
\includegraphics{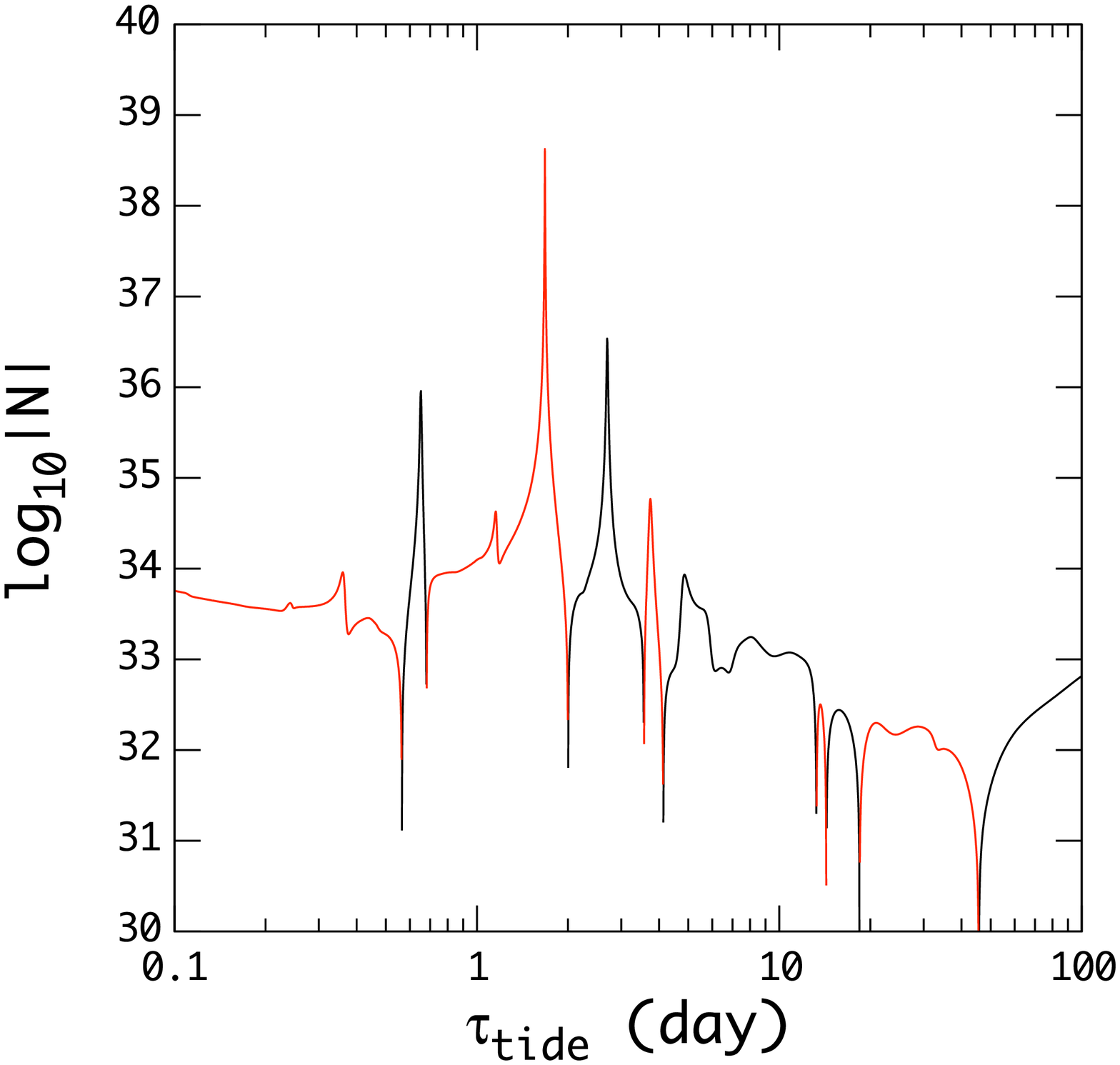}}
\caption{Tidal torque ${\cal N}$ versus
the tidal period $\tau_{\rm tide}=2\pi/\omega$ for $\bar\Omega_{\rm orb}=0.0537$ (left panel) and
for $\bar\Omega_{\rm orb}=-0.0537$ (right panel) where $\tau_*=1$ day and
${\rm Ek}=10^{-5}$ are assumed.
}
\label{fig:n_ekm5_peri_pmorb}
\end{figure}

Fig. \ref{fig:n_ekm5_peri_pmorb} plots ${\cal N}$ 
versus $\tau_{\rm tide}$ for $\bar\Omega_{\rm orb}=0.0537$ (left panel) and for $\bar\Omega_{\rm orb}=-0.0537$ (right panel)
where $\tau_*=1$ day and ${\rm Ek}=10^{-5}$ are assumed and the red (black) lines indicate positive (negative) ${\cal N}$.
Note that for $\bar\Omega_{\rm orb}>0$, positive (negative) ${\cal N}$ induces synchronization (asynchronization) 
between the orbital motion and spin of the planets and vice versa for $\bar\Omega_{\rm orb}<0$.
In Fig. \ref{fig:n_ekm5_peri_pmorb}, we find a series of sharp peaks of ${\cal N}$ caused by resonance with
the core inertial modes.
For $\bar\Omega_{\rm orb}=0.0537$, the most prominent peak of ${\cal N}$ occurs at the ratio $\omega/\Omega\approx 0.556$ ($\tau_{\rm tide}\approx9.38$ day),
which corresponds to the prograde $i_2$ inertial mode, and the second most prominent ones occur at $\omega/\Omega\approx 0.274$
($\tau_{\rm tide}\approx17.0$ day) and $\omega/\Omega\approx 1.27$
($\tau_{\rm tide}\approx 5.25$ day), corresponding to the prograde $i_4$ inertial modes (see Table 1).
The third peak in the series occurs at $\omega/\Omega\approx 0.163$ ($\tau_{\rm tide}\approx 27.1$ day), which may belong to $i_6$ inertial modes.
For $\bar\Omega_{\rm orb}=-0.0537$, on the other hand, the most prominent peak occurs at $\omega/\Omega\approx -1.10$
($\tau_{\rm tide}\approx1.68$ day), corresponding to the retrograde $i_2$ inertial mode and the second most prominent ones
at $\omega/\Omega\approx -0.862$ ($\tau_{\rm tide}\approx 2.69$ day) and at $\omega/\Omega\approx -1.52$ ($\tau_{\rm tide}\approx 0.653$ day), which are the retrograde $i_4$ inertial modes (see Table 1).
For $\bar\Omega_{\rm orb}=0.0537$, ${\cal N}$ is negative when $\tau_{\rm tide}\sim 100$ day,
and this suggests that  
the tide may hamper the process of synchronization when the system is close to complete synchronization.

For $\bar\Omega_{\rm orb}<0$, the forcing of $\tau_{\rm tide}>0$ is retrograde forcing and can tidally excite $r$-modes 
when $\tau_{\rm tide}\gtrsim 10$ day, corresponding to $\bar\omega\approx 2m\bar\Omega/l'(l'+1)\sim-\bar\Omega/3$. 
Since the $r$-modes of even parity propagate in the envelope in which radiative dissipation is significant, 
the resonance produces broad peaks of ${\cal N}$ as a function of $\tau_{\rm tide}$ as shown by the right panel of 
Fig. \ref{fig:n_ekm5_peri_pmorb}.
The resonance peaks due to the $r$-modes found in this paper, however, do not necessarily stand out clearly
compared to those computed by Lee \& Murakami (2019), who considered only thermal tides in the envelope.

\begin{figure}
\resizebox{0.45\columnwidth}{!}{
\includegraphics{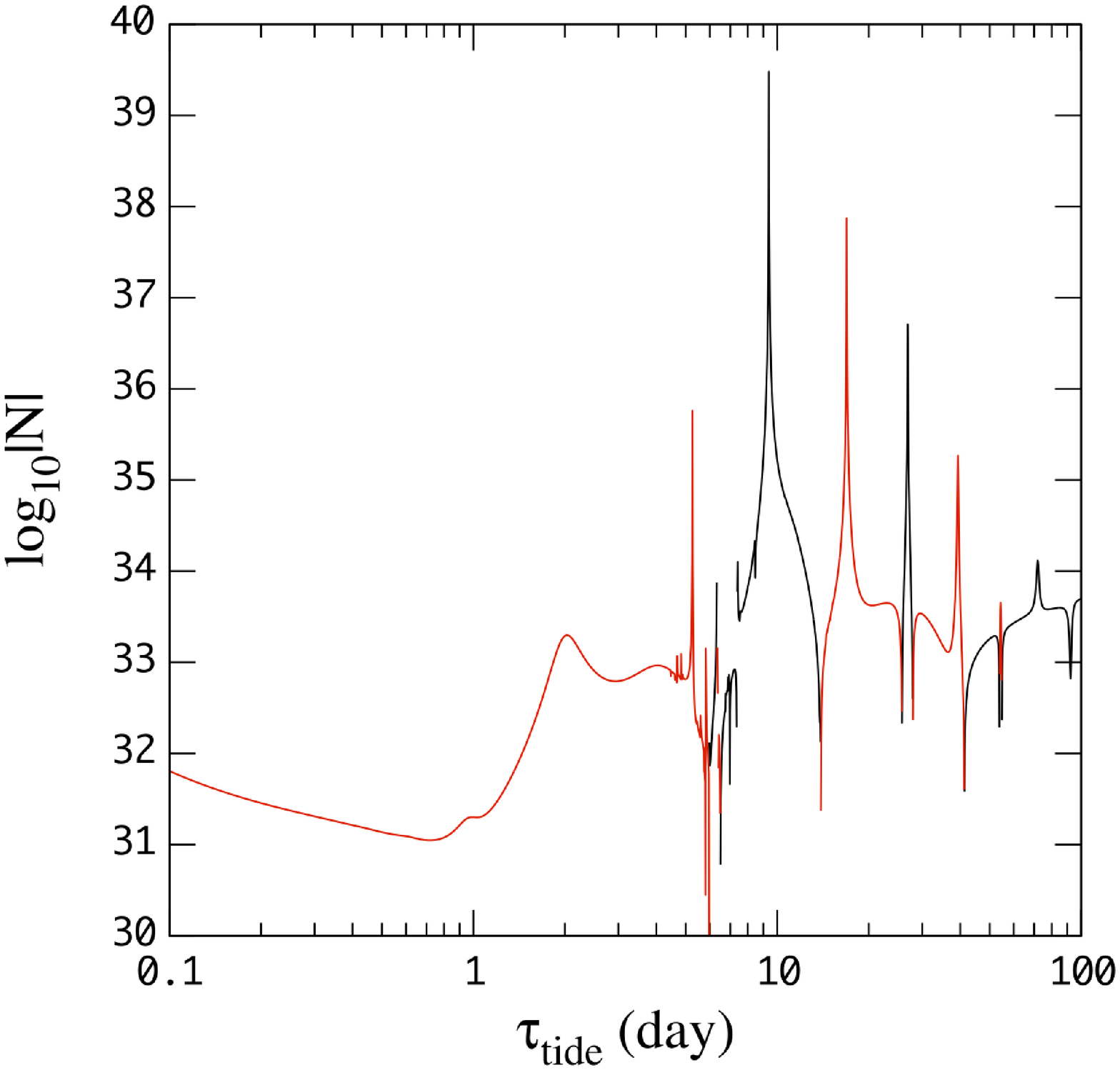}}
\hspace*{1cm}
\resizebox{0.45\columnwidth}{!}{
\includegraphics{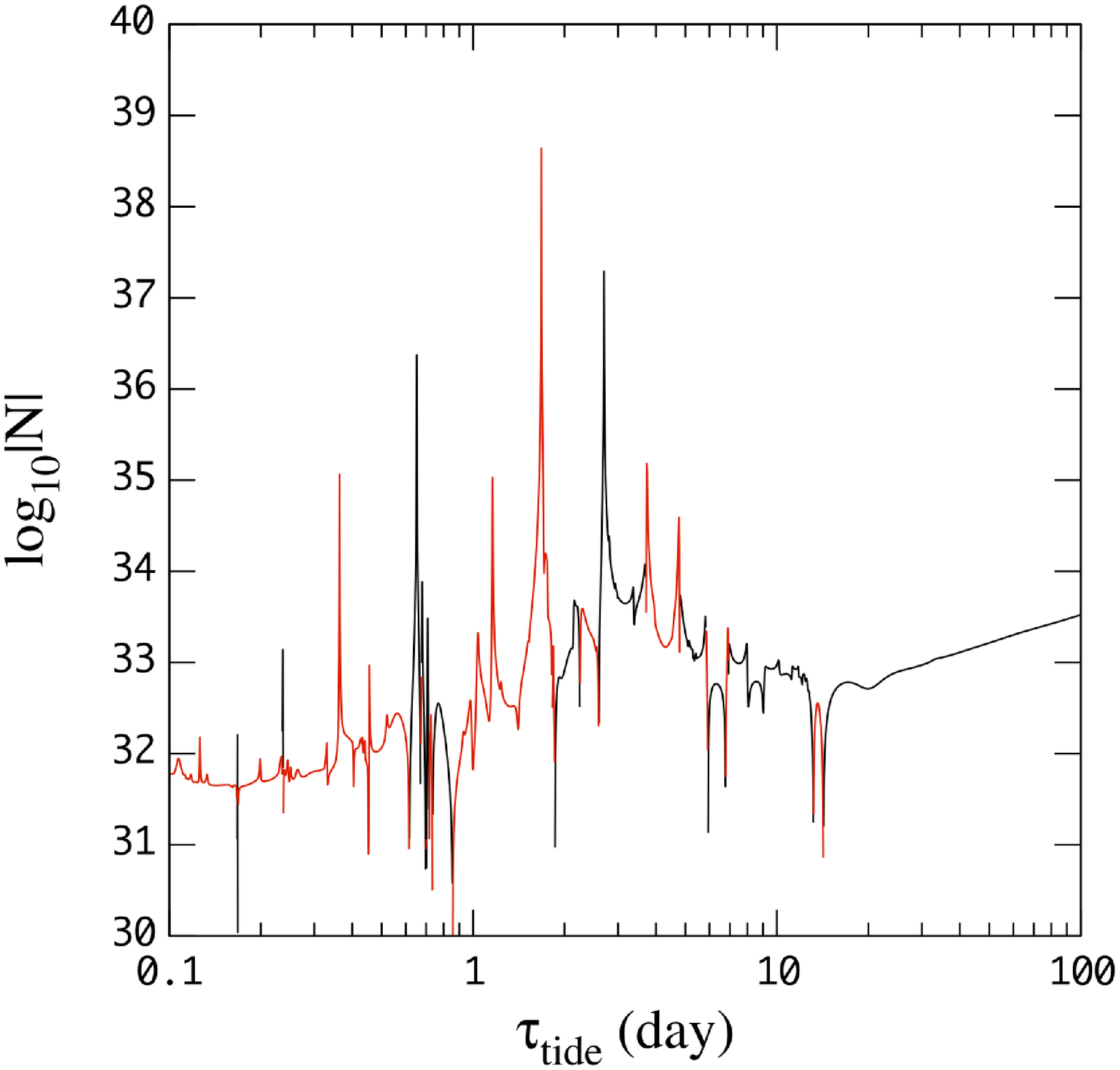}}
\caption{Same as Fig. \ref{fig:n_ekm5_peri_pmorb} but for ${\rm Ek}=10^{-7}$.
}
\label{fig:n_ekm7_peri_morb}
\end{figure}

Fig. \ref{fig:n_ekm7_peri_morb} plots ${\cal N}$ versus $\tau_{\rm tide}$ for ${\rm Ek}=10^{-7}$.
We find that the widths of the resonance peaks due to
the $i_2$ and $i_4$ inertial modes are narrower than those for ${\rm Ek}=10^{-5}$.
We also find that the number of minor resonance peaks that appear in the inertial range
is larger.
However, the gross properties of ${\cal N}$ as a function of $\tau_{\rm tide}$ are quite similar between
the two cases.

\subsubsection{Tidal Torque in the Interior}

\begin{figure}
\resizebox{0.45\columnwidth}{!}{
\includegraphics{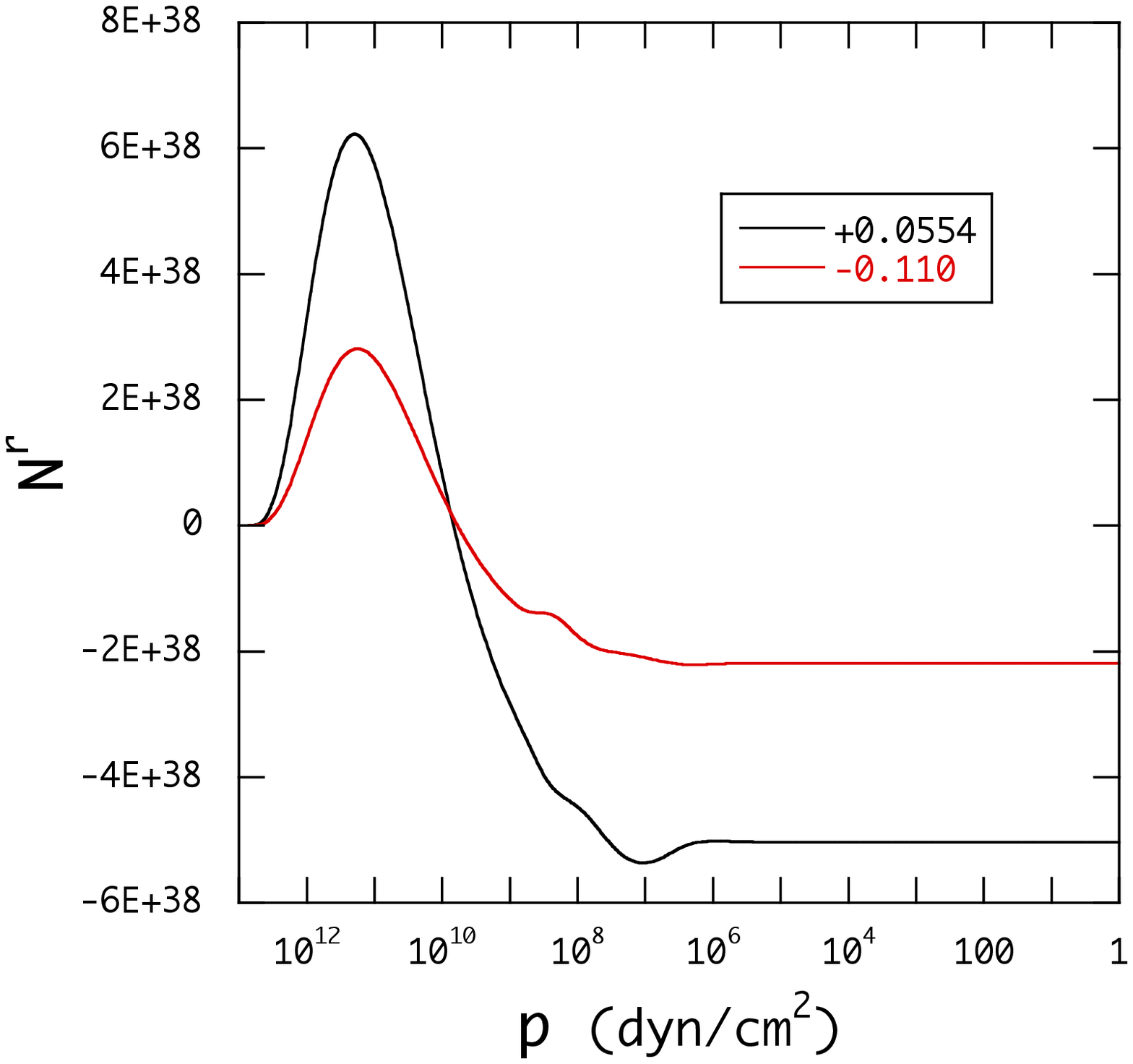}}
\hspace*{0.65cm}
\resizebox{0.45\columnwidth}{!}{
\includegraphics{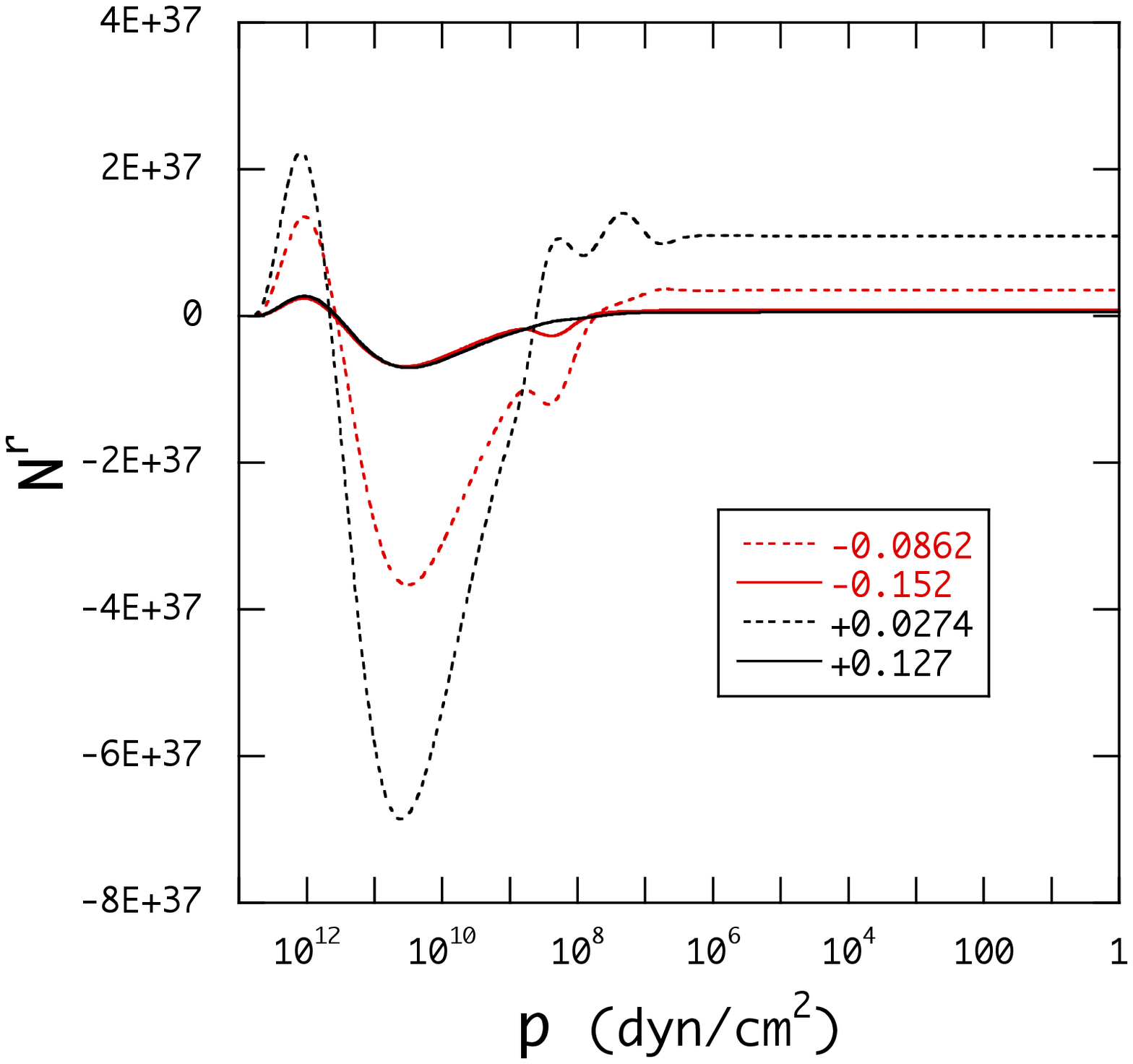}}
\caption{${\cal N}^r$ as a function of the gas pressure 
$p$ at the resonance with the $i_2$ (left panel) and $i_4$ (right panel) inertial modes where we assume
$\bar\Omega=0.1$, $\tau_*=1$ day, and 
Ek$=10^{-5}$.
The numbers in the insets indicate the normalized forcing frequency $\bar\omega$.
}
\label{fig:nri2i4}
\end{figure}

\begin{figure}
\resizebox{0.33\columnwidth}{!}{
\includegraphics{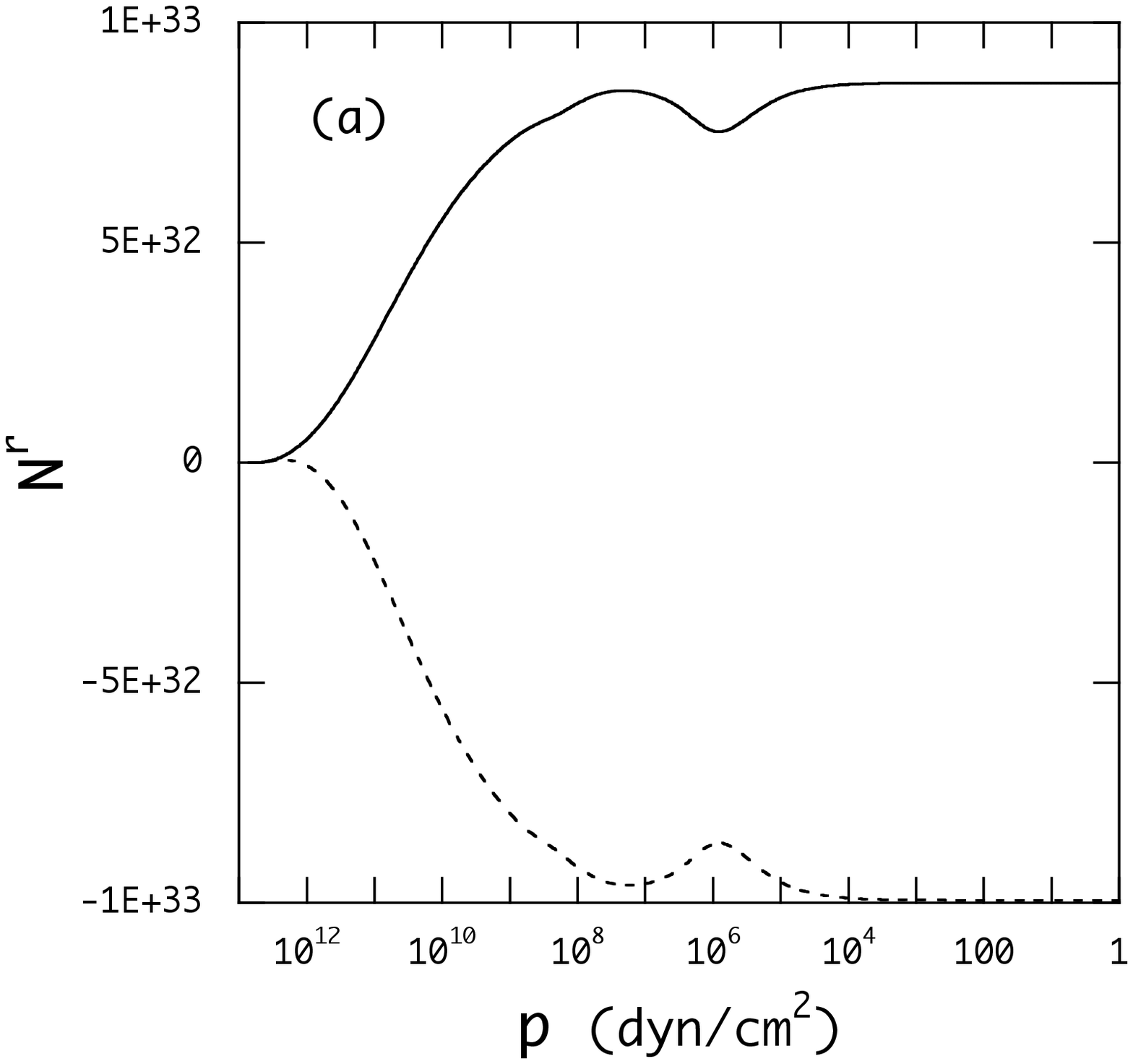}}
\resizebox{0.33\columnwidth}{!}{
\includegraphics{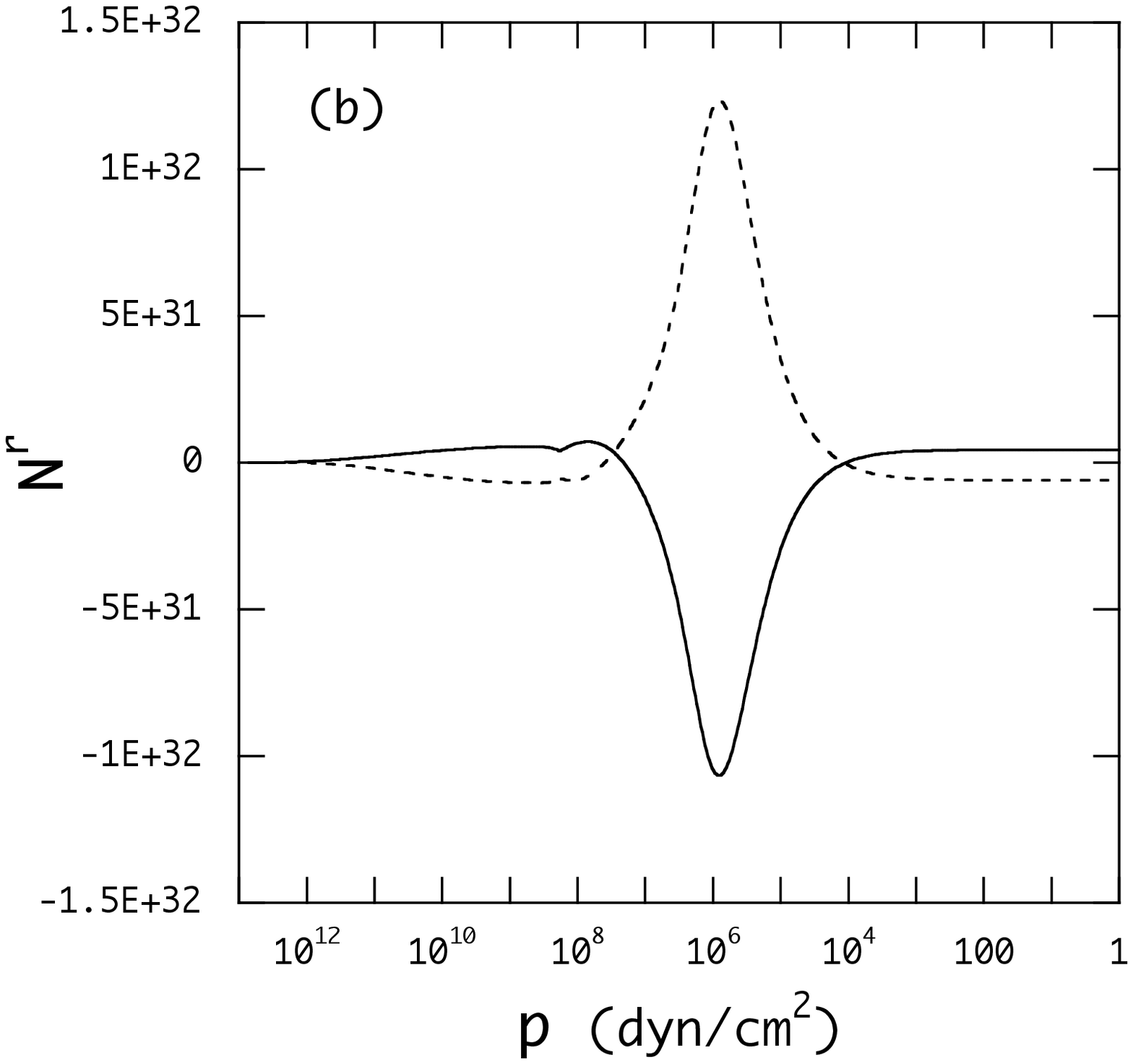}}
\resizebox{0.33\columnwidth}{!}{
\includegraphics{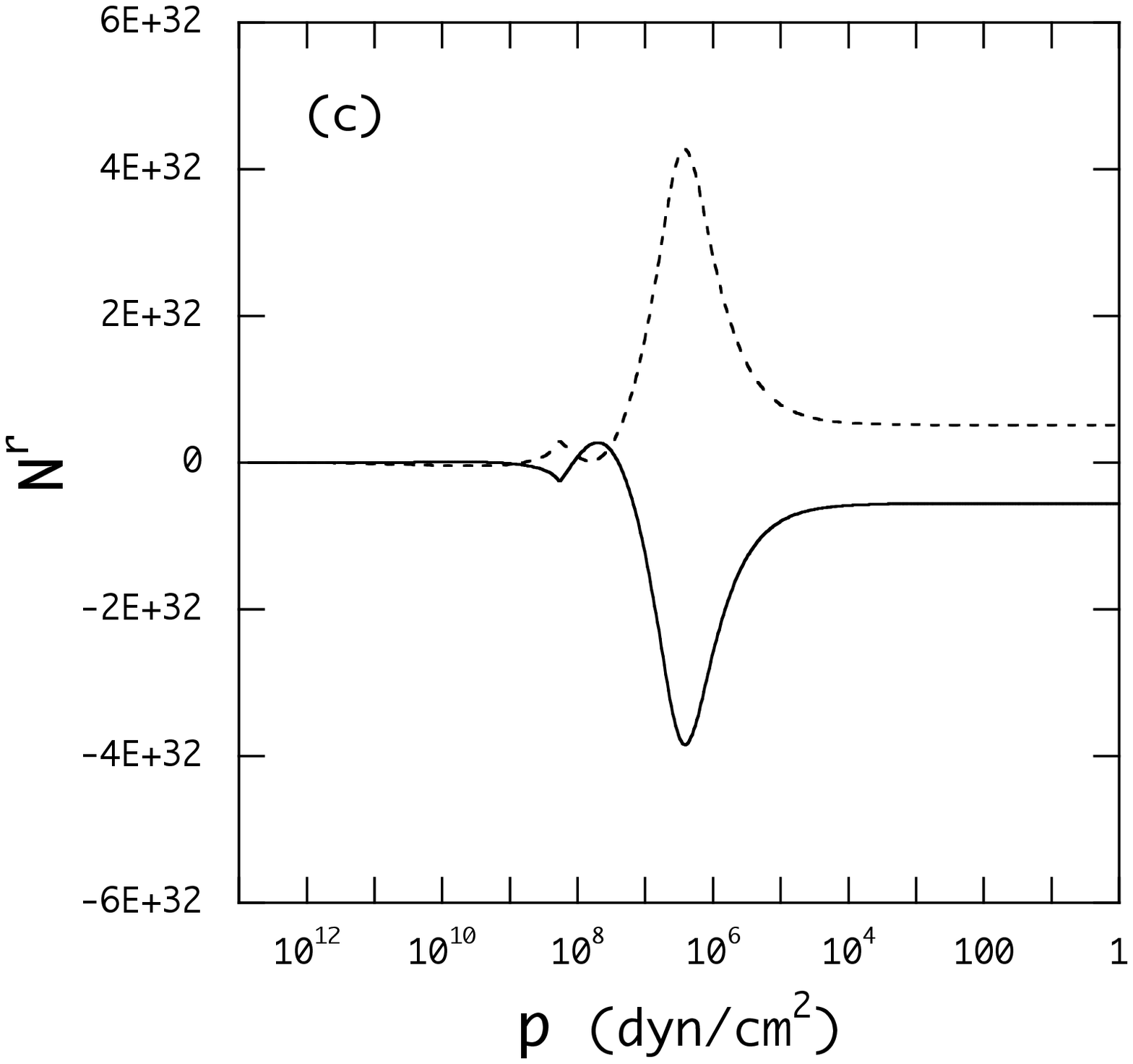}}
\caption{${\cal N}^r$ as a function of the gas pressure $p$ for the forcing frequencies $\bar\omega=0.4$ (solid lines) and
$\bar\omega=-0.4$ (dashed lines), where $(\tau_*,{\rm Ek})=(1,10^{-5})$ for panel (a), $(1,10^{-7})$ for panel (b), and
 $(0.1,10^{-7})$ for panel (c).
}
\label{fig:nr_pm04}
\end{figure}

Defining ${\cal N}^r(m,\omega)$ as
\be
{\cal N}^r(m,\omega)={m}\int_0^r{\rm Im}\left(\int\Phi_T\rho^{\prime *}\sin\theta d\theta d\phi \right)r^2 d r,
\ee
we plot ${\cal N}^r$ as a function of the gas pressure $p$ in Fig. \ref{fig:nri2i4} for the forcing frequencies
in resonance with the $i_2$ (left panel) and $i_4$ (right panel) inertial modes in the core.
At resonances with the core inertial modes, the amplitudes of the tidal responses are well confined in the core, and
thermal effects in the envelope have negligible contributions to ${\cal N}$, except for those from the regions around the
core-envelope boundary.
As the figure shows, viscous dissipation in the core has dominating effects on the tidal torques and
${\cal N}^r$ at the surface is negative for the $i_2$ inertial modes and
positive for the $i_4$ inertial modes.
If the forcing is off resonance, ${\cal N}^r$ at the surface may be determined by the sum of contributions
from the viscous core and radiative envelope as suggested by Fig. \ref{fig:nr_pm04}, which plots
${\cal N}^r$ at the forcing frequency $\bar\omega=\pm0.4$ for several combinations of $(\tau_*,{\rm Ek})$.
For Ek$=10^{-5}$ and $\tau_*=1$ day (panel a), the viscous contributions from the core dominates thermal contributions 
from the envelope and hence
${\cal N}^r$ at the surface is positive (negative) for prograde (retrograde) forcing, meaning that the tides drive the spin of the planets
to synchronize with the orbital motion.
This is still the case for Ek$=10^{-7}$ and $\tau_*=1$ day (panel b), although the viscous contributions from the core
are not necessarily dominating.
For Ek$=10^{-7}$ and $\tau_*=0.1$ day (panel c), however, the tidal responses in the outer envelope
are strongly non-adiabatic so that the changes in ${\cal N}^r$ there are smaller than those at the bottom
of the envelope and ${\cal N}^r$ at the surface becomes negative (positive) for prograde (retrograde) forcing.

\section{conclusions}

We have computed small amplitude gravitational and thermal tides in uniformly rotating hot Jupiters, composed of a thin radiative envelope and a nearly isentropic convective core.
In our numerical analyses,
we took account of the effects of radiative dissipation in the envelope and viscous dissipation
in the core on the tidal responses.
We used the approximation in which the radiative dissipation is given by $\nabla\cdot\pmb{F}'\sim\rho c_pT'/\tau_*$, where $\tau_*$ is a parameter for thermal heat exchange timescale.
Treating the convective fluids as viscous ones,
we solved linearized Navier-Stokes equation to obtain tidal responses of the core, where
the Ekman number Ek is assumed as a constant parameter for the viscosity coefficient in the rotating planets.
We matched tidal responses in the viscous core and in the radiative envelope at the core-envelope boundary 
with appropriate jump conditions to obtain a complete response.

We computed tidal torques ${\cal N}$
for rotating hot Jupiters.
Since the convective core is nearly isentropic and supports propagation of inertial modes, 
the behavior of the tidal responses depends on weather the tidal forcing $\omega$ is
in the inertial range or not, where the inertial range is defined as $|\omega/\Omega|\le 2$ for a given rotation rate $\Omega$.
For the tidal forcing outside the inertial range, we found that
the effects of viscous dissipations in the core 
are not always dominating when ${\rm Ek}\lesssim 10^{-7}$ in the sense that
viscous contributions to ${\cal N}$ in the core are comparable to or smaller than thermal contributions
in the radiative envelope.
In this case, the sign of ${\cal N}$ outside the inertial range is dependent on $\tau_*$ in the envelope.
However, for ${\rm Ek}\gtrsim 10^{-7}$, the sign of ${\cal N}$ does not depend on $\tau_*$ and is positive (negative)
for prograde (retrograde) forcing and the tides drive the planets to synchronization between the spin and orbital motion.
On the other hand, when the forcing $\omega$ is in the inertial range, 
frequency resonance between the forcing $\omega$ and inertial modes plays 
a significant role to determine ${\cal N}$.
If the tidal forcing is in resonance with low order inertial modes such as $i_2$ and $i_4$,
we obtain very high resonance peaks of ${\cal N}$.
We found that ${\cal N}$ in the resonance peaks is negative (positive) for the  
$i_2$ ($i_4$) inertial modes, and that the signs do not depend on parameter values of Ek and $\tau_*$.
As Ek decreases, however, frequency resonances with inertial modes $i_n$ with large $n$'s
will be important to determine ${\cal N}$ in the frequency regions between $i_2$ and $i_4$ inertial modes.

Using the tidal quality factor $Q$, 
the tidal torque due to equilibrium gravitational tides may be estimated as (e.g., Goldreich \& Soter 1966)
\be
{\cal N}_{\rm eq}={3GM_*^2R^5\over 2a_*^6}{1\over Q},
\ee
which is written as ${\cal N}_{\rm eq}=1.6\times10^{38}/Q$ for the parameter values used in this paper.
We note that $|{\cal N}|$ discussed in this paper strongly depends on the forcing frequency $\omega$.
If the forcing is outside the inertial range,
$|{\cal N}|$ is nearly proportional to Ek when ${\rm Ek}\gtrsim 10^{-7}$ and 
is of order of $10^{31}$ at $|\bar\omega|\sim 0.4$ for Ek$=10^{-7}$.
This value of ${\cal N}\sim10^{31}$ for Ek$=10^{-7}$ may correspond to
$Q\sim 10^7$, which is much larger than $Q\sim 10^5$ estimated for Jovian planets 
(e.g., Goldreich \& Soter 1966).
If the forcing is in the inertial range, on the other hand, it is difficult to 
define a mean value of $|{\cal N}|$ because numerous resonance peaks appear.
If we ignore these resonance peaks due to inertial modes, a mean value of $|{\cal N}|$ in the inertial range
may be determined by the thermal property and resonant $g$-modes in the envelope, particularly 
for ${\rm Ek}\lesssim 10^{-7}$.

{
It is interesting to compare the results obtained in this paper to those of previous investigations.
Savonije \& Papaloizou (1997), for example, computed fully non-adiabatic tidal responses of a uniformly rotating $20M_\odot$ 
main sequence star to low frequency tidal forcing, treating the fluid in the convective core as a viscous fluid.
They used a two dimensional difference code to compute oscillations of rotating stars.
The boundary conditions they used are similar to those used in this paper, except for jump conditions
at the core-envelope boundary.
They found that the spin-up or spin-down timescale, which is closely related to the tidal torque, strongly depends on
the forcing frequency and shows clear features characteristic to resonance with low frequency oscillation modes
such as $g$-modes and $r$-modes in the envelope.
They also showed an example of tidal responses for which the forcing at $\omega/\Omega\approx 0.37$ is in resonance with an inertial mode in the core.
We find that the responses such as $\xi_r$ in their paper for the inertial mode 
are quite similar to the expansion coefficient $xS_{l_1}$ of the $i_2$ mode of $\omega/\Omega\approx 0.55$.
The difference in the location $\omega/\Omega$ of the inertial mode
may be due to the difference in the structure of the convective core.
The structure of the core of a massive main sequence star is significantly different from that of a polytrope of the index $n=1$.
Another example may be given by Ogilvie \& Lin (2004), who computed tidal responses of rotating giant planets, 
composed of a solid core, a thick convective shell and a thin radiative envelope.
They treated the convective fluid in the shell as a viscous fluid, for which the Ekman number is introduced as a constant parameter for 
the viscous coefficient.
As a function of the forcing frequency, they computed the amount of viscous dissipations in the shell and of energy leakage from the envelope treating the responses in the envelope as traveling waves, instead of estimating thermal responses there.
To represents tidal responses of rotating planets, they expanded the perturbations using associated Legendre functions 
and derived a set of linear ordinary differential equations for the expansion coefficients, which they integrated 
using mainly a Chebyshev pseudospectral approach.
The set of linear differential equations should be almost the same as the set of equations solved in this paper
for the viscous fluids,
except that we include the effects of radiative dissipations on the responses. 
As a function of the forcing frequency, they found in the viscous dissipation rates in the inertial range 
clear resonance peaks due to the $i_2$ inertial modes and
that the number of minor resonance peaks increases with decreasing Ek, which results are quite similar to
those obtained in this paper.
Probably, the frequency spectrum of inertial modes they obtained is not necessarily the same as the spectrum
discussed in this paper because they assumed the presence of a solid core.

In a series of papers, Witte \& Savonije (1999, 2001, 2002) proposed a mechanism of locking the tidal forcing in
a state close to resonance with a low frequency oscillation mode of rotating stars in a binary system,
expecting that if the tidal locking lasts for a long period, the tides due to the resonance state
would strongly affect tidal evolution of the binary system.
They argued that if the eccentricity of the binary orbit is large enough there appears many possible 
forcing frequencies $n\Omega_{\rm orb}$ with integers $n$ and hence prograde and retrograde forcings in the co-rotating frame of the star and that if one of the forcing frequencies is close to resonance state with an oscillation mode,
there occurs a balance between the prograde and retrograde forcings and the forcings
would be locked in that balance state. 
As another tidal locking mechanism, 
we suggested in this paper that, if neighboring resonance peaks of the tidal torque due to core inertial modes
have alternatively different signs as a function of the forcing frequency or period, 
the process of synchronization would be hampered and the forcing
would be trapped between the two resonance peaks of different signs.
As a possible example for this mechanism, we suggested that since the sign of the resonance peak due to
the prograde $i_2$ inertial mode is negative and opposite to those of the neighboring peaks due to $i_4$ inertial modes,
the tidal forcing would be trapped between the resonance peaks due to the prograde $i_2$ and $i_4$ inertial modes.

}

{In this paper, we use the turbulent viscosity $\nu_{\rm turb}$ for the viscosity coefficient and estimate
Ekman number to be Ek$\sim10^{-7}$ in the convective core, using the values of $v_c$ and $l_m$ expected for Jupiter.
However, Goldreich \& Nicholson (1977) argued that if the tidal forcing period $\tau_T$ is shorter than the convective
turnover timescale $\tau_c$, the turbulent viscosity coefficient $\nu_{\rm turb}$ should be reduced by a factor $(\tau_T/\tau_c)^2$.
If this is the case, the reduction factor will be of order of $10^{-6}$ for a forcing frequency $\bar\omega\sim 0.1$ since we find $\tau_c\sim 3~{\rm year}$ deep in the convective core from Guillot et al (2004) for Jupiter and
$\tau_T\sim5\times 10^{-3}$ year for a Jovian planet.
This reduction factor reduces
Ek$\sim 10^{-7}$ used as the standard value in this paper to Ek$\sim 10^{-13}$.
If Ek$\sim 10^{-13}$ is assumed for the entire convective core, 
the viscous contributions to the tidal torque will be
negligible compared to the contributions from the radiative envelope in which radiative dissipations play
the dominant role as suggested by Fig. \ref{fig:nr_pm04}.
}

The results obtained in this paper depend on approximations we made to compute tidal responses.
The thermal properties of radiative envelope depend on the parameter $\tau_*$ and 
the nature of viscous dissipations in the core on prescriptions we employ for the viscous coefficients.
It is desirable to use more realistic hot Jupiters models obtained from evolution calculations of irradiated Jovian planets with appropriate equations of state and opacities, which makes it possible for us to carry out fully non-adiabatic computations of tidal oscillations.
Although we assumed constant Ekman numbers for the viscosity coefficients in rotating Jovian planets, it will be useful to try
different prescriptions for viscosity coefficients to compute dissipative tidal responses.
We note that the frequency spectrum of inertial modes discussed in this paper would be significantly modified if 
a solid core exists in the deep interior (e.g., Ogilvie \& Lin 2004).
Although we assumed $e=0$ for the tidal calculations, we need to extend our computation to
the cases of $e\not=0$ (e.g, Witte \& Savonije 2001) to investigate the roles of tidal interactions
played in the scenario of high-eccentricity migration of Jovian planets (e.g., Dawson \& Johnson 2018).
Discussions concerning non-linearity of the responses are definitely necessary to obtain reliable conclusions.

\section*{Acknowledgements}
The author greatly thanks the anonymous referee for pointing out errors in my formulation for
viscous oscillations.


\begin{appendix}

\section{Derivation of the oscillation equations for viscous core}

In this appendix, we show some details of derivation of the oscillation equations
for the viscous core.
The three components of the perturbed equation of motion (\ref{eq:linearizedeom}) may be given by
\be
-\rho\omega^2\xi_r-2\rmi \omega\Omega\rho\xi_\phi\sin\theta=-{\partial p^\prime\over\partial r}
-\rho'{d\Phi\over dr}-\rho{\partial\Phi_T\over\partial r}
+{1\over r^2}{\partial\over\partial r}r^2\sigma_{rr}'+{1\over r\sin\theta}{\partial\over\partial\theta}\sin\theta\sigma_{r\theta}'
+{1\over r\sin\theta}{\partial\over\partial\phi}\sigma_{r\phi}'-{\sigma_{\theta\theta}'+\sigma_{\phi\phi}'\over r},
\label{eq:eom_r}
\ee
\be
-\omega^2\rho\xi_\theta-2\rmi \omega\Omega\xi_\phi\cos\theta=-{1\over r}{\partial p'\over \partial\theta}-\rho{1\over r}{\partial \Phi_T\over \partial\theta}+{1\over r^2}{\partial\over\partial r}r^2\sigma_{r\theta}'+{1\over r\sin\theta}{\partial\over\partial\theta}\sin\theta\sigma_{\theta\theta}'+{1\over r\sin\theta}{\partial\over\partial\phi}\sigma_{\theta\phi}'+{\sigma_{r\theta}'\over r}-{\sigma_{\phi\phi}'\over r}\cot\theta,
\label{eq:eom_theta}
\ee
\be
-\omega^2\rho\xi_\phi+2\rmi\omega\Omega\rho\left(\xi_\theta\cos\theta+\xi_r\sin\theta\right)&=&-{1\over r\sin\theta}{\partial p'\over\partial\phi}-\rho{1\over r\sin\theta}{\partial \Phi_T\over\partial\phi}
+{1\over r^2}{\partial\over\partial r}r^2\sigma_{r\phi}'+{1\over r\sin\theta}{\partial\over\partial\theta}\sin\theta\sigma_{\theta\phi}'\nonumber\\
&&+{1\over r\sin\theta}{\partial\over\partial\phi}\sigma_{\phi\phi}'+{\sigma_{r\phi}'\over r}+{\sigma_{\phi\theta}'\over r}\cot\theta,
\label{eq:eom_phi}
\ee
where $\sigma'_{ij}$ may be obtained by replacing $\pmb{v}$ in equations (6) to (11) with $\pmb{v}^\prime=\rmi\omega\pmb{\xi}$.
Substituting the expansions given by (\ref{eq:xiexp_r}), (\ref{eq:xiexp_theta}), (\ref{eq:xiexp_phi}),
and (\ref{eq:pexp}), we obtain the radial component of the equation of motion (\ref{eq:eom_r}) given as
\begin{eqnarray}
-c_1\bar\omega^2\sum_lS_lY_l^m-2\rmi c_1\bar\omega\bar\Omega\left(\sum_l\rmi mH_lY_l^m-\sum_{l'}T_{l'}\sin\theta{\partial\over\partial\theta} Y_{l'}^m\right)
= \sum_l{P_l\over \rho g}Y_l^m,
\label{eq:eom2_r}
\end{eqnarray}
where 
\be
P_l=-\rho gr{\partial \over\partial r}Z_{2,l}
-\rho g{d\ln\rho gr\over d\ln r}Z_{2,l}
+\rho g{\Phi_{T,l}\over gr}{d\ln \rho\over d\ln r}-\rho g{\rho'_l\over\rho}
+{2\rmi \omega\eta\over r}\left(2r{\partial S_l\over\partial r}+\Lambda_l H_l\right)-{\rmi\omega\eta\over r}\Lambda_lS_l-{\rmi\omega\eta\over r}\Lambda_lr{\partial H_l\over\partial r},
\ee
\be
Z_{2,l}=z_{2,l}+{\Phi_{T}^l\over gr},
\ee
where the variable $z_{2,l}$ is defined as
\be
z_{2,l}={p_l'\over\rho gr}-{p\over\rho gr}\left[\left({\zeta\over p}-{2\over 3}{\eta\over p}\right)\left({1\over r^2}{\partial\over \partial r}r^3\rmi\omega S_l-\rmi\omega \Lambda_lH_l\right)+2{\eta\over p}{\partial\over\partial r}\rmi\omega rS_l\right],
\label{eq:z2l}
\ee
and $\Lambda_l=l(l+1)$ and $g=GM_r/r^2$ with $M_r=\int_0^r4\pi r^2\rho dr$.
Similarly, substituting the expansions into equations (\ref{eq:eom_theta}) and (\ref{eq:eom_phi}), we
obtain
\be
-\omega^2\rho\xi_\theta-2\rmi\omega\Omega\rho\xi_\phi\cos\theta=\sum_lQ_l{\partial \over\partial\theta}Y_l^m(\theta,\phi)+
\sum_{l'}R_{l'}{1\over \sin\theta}{\partial\over\partial\phi}Y_{l'}^m(\theta,\phi),
\label{eq:eom2_theta}
\ee
\be
-\omega^2\rho\xi_\phi+2\rmi\omega\Omega\rho\left(\xi_\theta\cos\theta+\xi_r\sin\theta\right)=\sum_lQ_l{1\over \sin\theta}{\partial\over\partial\phi}Y_l^m(\theta,\phi)-\sum_{l'}R_{l'}{\partial\over\partial\theta}Y_{l'}^m(\theta,\phi),
\label{eq:eom2_phi}
\ee
where
\begin{eqnarray}
Q_l
=-\rho gr{Z_{2,l}\over r}-2{\rmi\omega\eta\over r}r{\partial\over\partial r}S_l+{1\over r^3}{\partial\over\partial r}\left[r^4{\rmi\omega\eta\over r}\left(r{\partial\over\partial r}H_l+S_l\right)\right]-2{\rmi\omega\eta\over r}\Lambda_lH_l+2{\rmi\omega\eta\over r}H_l,
\end{eqnarray}
\be
R_{l'}={1\over r^3}{\partial\over\partial r}\left(r^4{\rmi\omega\eta\over r}r{\partial\over\partial r}T_{l'}\right)
-{\rmi\omega\eta\over r}\Lambda_{l'}T_{l'}+2{\rmi\omega\eta\over r}T_{l'}.
\ee
It is convenient to use the divergence given by $\sin^{-1}\theta\partial_\theta\sin\theta({\rm eq.}~\ref{eq:eom2_theta})
+\sin^{-1}\theta\partial_\phi({\rm eq.}~\ref{eq:eom2_phi})$ and the radial component of curl, that is,
$\sin^{-1}\theta\partial_\theta\sin\theta({\rm eq.}~\ref{eq:eom2_phi})-\sin^{-1}\theta\partial_\phi({\rm eq.}~\ref{eq:eom2_theta})$, which reduce to
\begin{eqnarray}
&&c_1\bar\omega^2\sum_l\Lambda_lH_lY_l^m -2\rmi c_1\bar\omega\bar\Omega \sum_{l'}\Lambda_{l'}T_{l'}\cos\theta Y_{l'}^m
+2\rmi c_1\bar\omega\bar\Omega\left(\sum_l\rmi mH_lY_l^m-\sum_{l'}T_{l'}\sin\theta\partial_\theta Y_{l'}^m\right)\nonumber\\
&&-2mc_1\bar\omega\bar\Omega\sum_lS_lY_l^m=-\sum_l\Lambda_l{Q_l\over\rho g}Y_l^m,
\label{eq:dz3}
\end{eqnarray}
and
\begin{eqnarray}
&&c_1\bar\omega^2\sum_{l'}\Lambda_{l'}T_{l'}Y_{l'}^m+2\rmi c_1\bar\omega\bar\Omega\sum_l\Lambda_lH_l\cos\theta Y_l^m
+2\rmi c_1\bar\omega\bar\Omega\left(\sum_lH_l\sin\theta\partial_\theta Y_l^m+\sum_{l'} m\rmi T_{l'}Y_{l'}\right)\nonumber\\
&&-2\rmi c_1\bar\omega\bar\Omega\sum_l\left(S_l\sin\theta\partial_\theta Y_l^m+2\cos\theta S_lY_l^m\right)
=\sum_{l'}\Lambda_{l'}{R_{l'}\over \rho g}Y_{l'}^m,
\label{eq:dz5}
\end{eqnarray}
where
$
\partial_\theta={\partial/\partial\theta}
$
and
$\partial_\phi={\partial/\partial\phi}.
$

Since we use as an independent variable the variable $z_{2,l}$ instead of $p'_l/\rho gr$, we have to 
represent $\rho'_l/\rho$ in terms of $z_{2,l}$.
Making use of equation (\ref{eq:z2l}),
we obtain for the density perturbation $\rho'_l$
\be
{\rho'_l\over\rho}
=\left(-rA-\beta_1+\rmi{3\alpha_2\over\hat\Gamma_1}\right)S_l+{V\over\hat\Gamma_1}z_{2,l}
+\rmi{2\alpha_1/3-\alpha_2\over\hat\Gamma_1}\Lambda_lH_l+\rmi{4\alpha_1/3+\alpha_2\over\hat\Gamma_1}r{\partial S_l\over\partial r}-{\alpha_T\over\rmi\omega+\omega_D}{\epsilon'_l\over Tc_p}.
\ee
Eliminating $\rho'_l$ from the linearized continuity equation given by
\be
\rho'_l+{1\over r^2}{\partial\over\partial r}r^3\rho S_l-\rho \Lambda_lH_l=0,
\ee
we obtain
\be
\left(1+\rmi{\alpha_2+4\alpha_1/3\over\hat\Gamma_1}\right)r{\partial\over\partial r}S_l
=\left({V\over\hat\Gamma_1}-3-3\rmi{\alpha_2\over\hat\Gamma_1}\right)S_l-{V\over\hat\Gamma_1}z_{2,l}
+\left(1+\rmi{\alpha_2-2\alpha_1/3\over\hat\Gamma_1}\right)\Lambda_lH_l
+{\alpha_T\over\rmi\omega+\omega_D}{\epsilon'_l\over Tc_p}.
\label{eq:dz1}
\ee

We derive the oscillation equations given in \S 2.2.2 for the viscous core from equations (\ref{eq:dz1}), (\ref{eq:eom2_r}),
(\ref{eq:dz3}), and (\ref{eq:dz5}), making use of the relations given by
\be
\sin\theta{\partial Y_l^m\over\partial\theta}=lJ_{l+1}^mY_{l+1}^m-(l+1)J_l^mY_{l-1}^m,
\ee
\be
\cos\theta Y_l^m=J_{l+1}^mY_{l+1}^m+J_l^mY_{l-1}^m.
\ee

\end{appendix}


\begin{thebibliography}{}


\bibitem[Arras P., Socrates A.(2010)]{2010ApJ..714..1} Arras P., Socrates A., 2010, ApJ, 714, 1

\bibitem[Auclair-Desrotour P., Leconte J.(2018)]{2018A&A..613..A45} Auclair-Desrotour P., Leconte J., 2018, A\&A, 613, A45

\bibitem[Baraffe I.etal(2003)]{2003A&A..402..701} Baraffe I., Chabrier G., Barman T.S., Allard F., Hauschildt P.H., 2003, A\&A, 402, 701


\bibitem[Bodenheimer etal. (2001)]{2001ApJ..548..466} Bodenheimer P., Lin D.N.C., Mardling R.A., 2001, ApJ, 548, 466

\bibitem[Dawson R.I., Johnson J.A.]{2018, Annu. Rev. Astron. Astrophys.} Dawson R.I., Johnson J.A., 2018, Annu. Rev. Astron. Astrophys., 56, 175

\bibitem[Eggleton P.P., Kiseleva L.G., Hut P.]{1998, ApJ, 499, 853} Eggleton P.P., Kiseleva L.G., Hut P., 1998, ApJ, 499, 853

\bibitem[Eggleton P.P., Kiseleva-Eggleton L.]{2001, ApJ, 562, 1012} Eggleton P.P., Kiseleva-Eggleton L., 2001, ApJ, 562, 1012

\bibitem[Fuller J., Lai D.]{2013MNRAS..430..274} Fuller J., Lai D., 2013, MNRAS, 430, 274

\bibitem[Goldreich P., Nicholson P.D.]{1977Icarus..30..301} Goldreich P., Nicholson P.D., 1977, Icarus, 30, 301

\bibitem[Goldreich, Soter]{Icarus1966..5..375} Goldreich P., Soter S., 1966, Icarus, 5, 375

\bibitem[Goldreich P., Keeley D.A.]{1977, ApJ, 211, 934} Goldreich P., Keeley D.A., 1977, ApJ, 211, 934

\bibitem[Guillot T. etal]{Jupiter. The planet, satellites and magnetosphere.} Guillot T., Stevenson D.J., Hubbard W.B., Saumon D., 2004, in Jupiter. The planet, satellites and magnetosphere. Eds. F. Bagenal, T.E. Dowling, W.B. McKinnon. Cambridge planetary science, Vol. 1, Cambridge University Press, 
Cambridge, UK

\bibitem[Guillot T.]{2005, Ann. Rev. Earth Planet. Sci, 33, 493} Guillot T., 2005, Ann. Rev. Earth Planet. Sci, 33, 493

\bibitem[Iro N.etal(2005)]{2005A&A..436..719} Iro N., B\'ezard B., Guillot T., 2005, A\&A, 436, 719


\bibitem[Jermyn (2107)]{2017MNRAS..469...1768} Jermyn A.D., Tout C.A., Ogilvie G.I., 2017, MNRAS, 469, 1768 


\bibitem[Landau L.D., Lifshitz E.M.]{1987, Fluid Mechanics, 2nd ed.} Landau L.D., Lifshitz E.M., 1987, Fluid Mechanics, 2nd ed., Pergamon Press, Oxford

\bibitem[Lee U.]{2019, MNRAS, 484, 5845} Lee U., 2019, MNRAS, 484, 5845

\bibitem[Lee U., Murakami D.]{2019, MNRAS, 488, 1960} Lee U., Murakami D., 2019, MNRAS, 488, 1960

\bibitem[Lee U., Saio H.(1986)]{1986MNRAS..221..365} Lee U., Saio H., 1986, MNRAS, 221, 365

\bibitem[Lee U., Saio H.(1987)]{1987MNRAS..224..513} Lee U., Saio H., 1987, MNRAS, 224, 513

\bibitem[Lee U., Saio H.(1990b)]{1990bApJ..360..590} Lee U., Saio H., 1990, ApJ, 360, 590

\bibitem[Lee U., Saio H.(1997)]{1997ApJ..491..839} Lee U., Saio H., 1997, ApJ, 491, 839

\bibitem[Lockitch K.H., Friedman J.L.]{1999, ApJ, 521, 764} Lockitch K.H., Friedman J.L., 1999, ApJ, 521, 764

\bibitem[Ogilvie G.I.]{2014Annu.Rev.Astron.Astrophys..52..171} Ogilvie G.I., 2014, Annu. Rev. Astron. Astrophys., 52, 171

\bibitem[Ogilvie G.I., Lin N.D.C.]{2004ApJ..610..477} Ogilvie G.I., Lin N.D.C., 2004, ApJ, 610, 477

\bibitem[Papaloizou J.C.B., Ivanov P.B.]{2010, MNRAS, 497, 1631} Papaloizou J.C.B., Ivanov P.B., 2010, MNRAS, 497, 1631

\bibitem[Press W.H., Teukolsky S.A.]{1977ApJ..213..183} Press W.H., Teukolsky S.A., 1977, ApJ, 213, 183

\bibitem[Savonije G.J., Papaloizou J.C.B.]{1984MNRAS..207..685} Savonije G.J., Papaloizou J.C.B., 1984, MNRAS, 207, 685
\bibitem[Savonije G.J., Papaloizou J.C.B.]{1997MNRAS..291..633} Savonije G.J., Papaloizou J.C.B., 1997, MNRAS, 291, 633

\bibitem[Savonije G.J., Witte M.G.]{2002, A&A, 386, 211} Savonije G.J., Witte M.G., 2002, A\&A, 386, 211

\bibitem[Socrates A.]{2013, arXiv:1304.4121} Socrates A., 2013, arXiv:1304.4121
  
\bibitem[Stevenson D.J.(1979)]{Geophys.Astrophy.Fluid.Dynamics1979..12..139} Stevenson D.J., Geophys. Astrophy. Fluid. Dynamics, 1979, 12, 139

\bibitem[Stevenson D.J.(1977a)]{1977ApJS..35..221} Stevenson D.J., Salpeter E.E., 1977a, ApJS, 35, 221

\bibitem[Stevenson D.J.(1977b)]{1977ApJS..35..239} Stevenson D.J., Salpeter E.E., 1977b, ApJS, 35, 239

\bibitem[Stevenson D.J.]{1982, Planet. Space Sci., 30, 755} Stevenson D.J., 1982, Planet. Space Sci., 30, 755 

\bibitem[Terquem C., Papaloizou J.C.B., Nelson R.P., Lin D.N.C.]{1998, ApJ, 502, 788} Terquem C., Papaloizou J.C.B., Nelson R.P., Lin D.N.C., 1998, ApJ, 502, 788

\bibitem[Udry S., Santos N.C.]{2007, Annu. Rev. Astron. Astrophys., 45, 397} Udry S., Santos N.C., 2007, Annu. Rev. Astron. Astrophys., 45, 397 

\bibitem[Witte M.G., Savonije G.J.]{1999AA..350..129} Witte M.G., Savonije G.J., 1999, A\&A, 350, 129

\bibitem[Witte M.G., Savonije G.J.]{2001AA..366..840} Witte M.G., Savonije G.J., 2001, A\&A, 366, 840

\bibitem[Witte M.G., Savonije G.J.]{2002AA..386..222} Witte M.G., Savonije G.J., 2002, A\&A, 386, 222

\bibitem[Wu Y.]{2005, ApJ, 635, 688} Wu Y., 2005, ApJ, 635, 688

\bibitem[Wu Y., Murray N.]{2003, ApJ, 589, 605} Wu Y., Murray N., 2003, ApJ, 589, 605

\bibitem[Xiong D.R., Cheng Q.L., Deng L.]{1997, ApJS, 108, 529} Xiong D.R., Cheng Q.L., Deng L., 1997, ApJS, 108, 529


\bibitem[Yoshida S., Lee U. (2000)]{2000ApJ..529..997} Yoshida S., Lee U., 2000, ApJ, 529, 997

\bibitem[Zahn J.P.]{1966, AnAp, 29, 489} Zahn J.P., 1966, AnAp, 29, 489


\bibitem[Zahn J.P.]{1989, AA, 220, 112} Zahn J.P., 1989, A\&A, 220, 112



\end{thebibliography}


\end{document}